\documentclass[aps,article,twocolumn,superscriptaddress]{revtex4-2}

\usepackage[utf8]{inputenc}
\usepackage[fleqn]{amsmath}
\usepackage{amssymb}
\usepackage{graphicx}
\usepackage{ulem,xcolor}
\usepackage{epstopdf, epsfig}
\usepackage{graphicx}
\usepackage{mathtools}
\usepackage{bm}

\newcommand{\Wcs}{W}
\newcommand{\Wbw}{\overline{W}}

\begin{document}

\title{Kinetic structure of strong-field QED showers in crossed electromagnetic fields}

\author{M. Pouyez}
\affiliation{LULI, Sorbonne Université, CNRS, CEA, École Polytechnique, Institut Polytechnique de Paris, F-75255 Paris, France}
\author{T. Grismayer}
\affiliation{GoLP/Instituto de Plasmas e Fusão Nuclear, Instituto Superior Técnico, Universidade de Lisboa, 1049-001 Lisboa, Portugal}
\author{M. Grech}
\affiliation{LULI, CNRS, CEA, Sorbonne Universit\'{e}, École Polytechnique, Institut Polytechnique de Paris, F-91128 Palaiseau, France}
\author{C. Riconda}
\affiliation{LULI, Sorbonne Université, CNRS, CEA, École Polytechnique, Institut Polytechnique de Paris, F-75255 Paris, France}

\begin{abstract}
    A complete, kinetic description of electron-seeded strong-field QED showers in crossed electromagnetic fields is derived. The kinetic structure of the shower and its temporal evolution are shown to be a function of two parameters: the initial shower quantum parameter and radiation time. The latter determines the short and long time evolution of the shower. Explicit solutions for the shower multiplicity (number of pairs per seed electron) and the emitted photon spectrum are obtained for both timescales. Our approach is first derived considering showers in a constant, homogeneous magnetic field. We find that our results are valid for any crossed fields and we apply them to laboratory settings for which we obtain fully analytical, predictive scaling laws.
\end{abstract}

\maketitle

%\section{Introduction}
%_________________________________________________________________%

Electromagnetic showers (EMS) are observed when a high-energy particle, such as an electron or a photon, interacts with matter. These showers are characterized by a cascade of secondary particles, primarily electrons, positrons, and photons, that result from a series of electromagnetic (EM) interactions, and play a crucial role in particle physics and astrophysics. EMS were first conjectured and studied with the Bremstrahlunng and the Bethe-Heitler process \cite{bhabha1937passage,carlson1937multiplicative,landau1938cascade}. The article of Landau \cite{landau1938cascade} was the first to provide a rigorous method to compute the number of particles as a function of penetration depth
as well as the energy distribution for shower particles at a given depth. 
Today EMS are the main candidate to produce quasi-neutral pair beams using high-intensity lasers \cite{HChen2023}, wakefield electron beams \cite{generation_sarri_2015}, and accelerator proton beams \cite{Arrowsmith_2024}. High multiplicity is necessary to obtain quasi-neutral pair beams and it can only be obtained with kJ lasers \cite{HChen2023} or ultra-relativistic particle beams \cite{Arrowsmith_2024}.
With several PW laser facilities being functional worldwide, 
strong-field quantum electrodynamics (SF-QED) has become a hot research topic \cite{di2012extremely,gonoskov2022charged,fedotov2023advances}. In an ultra-intense electromagnetic field, processes like nonlinear inverse Compton scattering (nICS) and multiphoton Breit-Wheeler (nBW) play the analogous role of Bremsstrahlung and Bethe-Heitler. SF-QED EMS were first speculated to play an important role in a Neutron star’s polar cap \cite{sturrock1971model,ruderman1975theory} for coherent emission, which led to several theoretical works to estimate the pair multiplicity in uniform \cite{akhiezer1994kinetic} and curved \cite{hibschman2001pair1,hibschman2001pair2} magnetic fields. Akhiezer et al. \cite{akhiezer1994kinetic} tried to apply Landau’s method which was proven inadequate by Monte Carlo simulations of the SF-QED process  \cite{anguelov1999electromagnetic}.
A generation-splitting method was used in  \cite{hibschman2001pair1,hibschman2001pair2} to integrate the kinetic equations, obtaining
estimates of the pair multiplicity for pulsar parameters and geometry. More recently SF-QED EMS have been investigated in the context of laser-electron scattering in the weak quantum regime ($\chi \lesssim 1$) \cite{blackburn2017scaling,mercuri2021impact} and the medium quantum regime ($\chi < 50$) \cite{lobet2017generation,qu2022collective,pouyez2024multiplicity,matheron2024self}, the later works showing that high-multiplicity can be reached with multi-PW lasers. Here $\chi$ is the so-called quantum parameter \cite{SuppMat}.

Although advanced numerical tools to explore showers are readily available, a complete theory of SF-QED showers in all quantum regimes is yet to be found and no explicit solution is proposed in the literature. It is 
essential to have reliable analytical results and scaling laws that offer a detailed understanding of this fundamental QED process. In this Letter, we use the generation-splitting method to investigate in depth the kinetic structure and time-evolution of SF-QED showers. We derive explicitly the shower multiplicity, a key quantity for the production of quasi-neutral pair plasmas in the laboratory. We show that the problem is only a function of two parameters: the initial shower quantum parameter $\chi_0$ and radiation time $T_r$. The latter delimits the short and long time evolution of the shower which correspond to two different regimes. The maximum number of generations participating to the process is also found. This analysis being valid for any crossed EM fields, we conclude with applications of our results to two different laboratory settings.

\begin{figure*}[htbp]
    \centering
    \includegraphics[width=\textwidth]{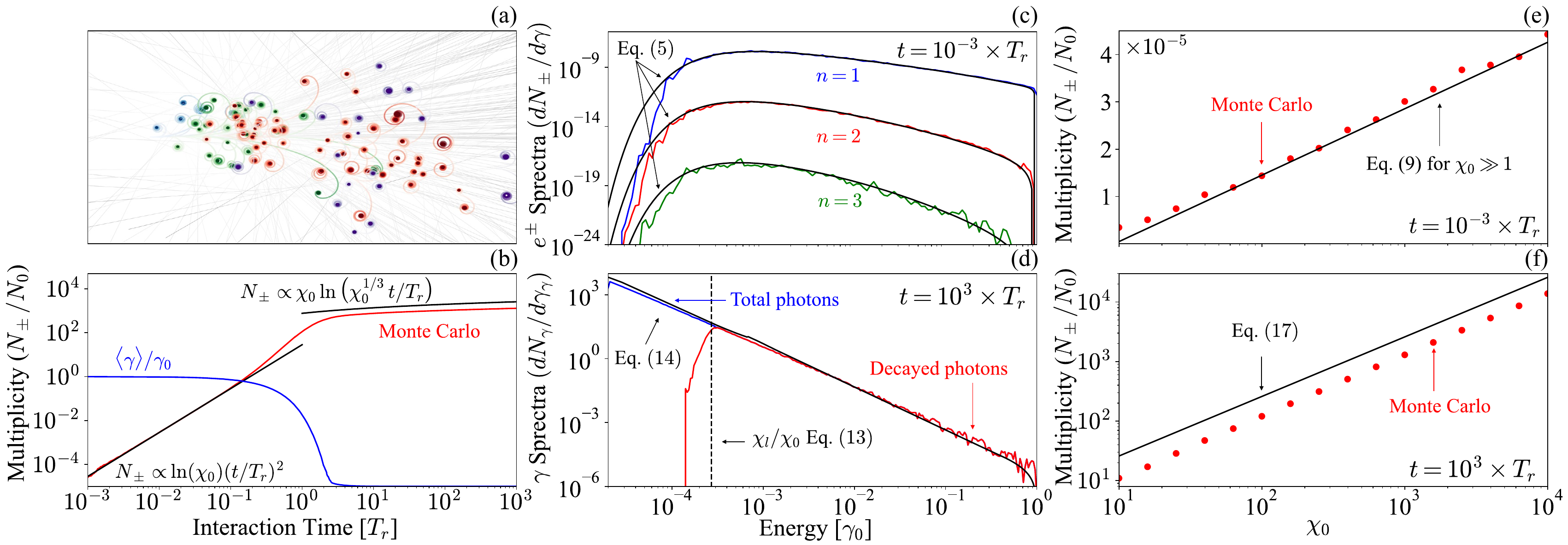} 
    \caption{Electron-seeded SF-QED shower in a constant uniform magnetic field. (a) Particle trajectories in the plane perpendicular to the magnetic field for an EMS seeded $5$ seeded electrons. Leptons are shown in color, each color corresponding to a generation. Photon trajectories are in grey. (b) Temporal evolution of the shower multiplicity (red) and average seed-electron energy (blue) from MC simulations. The time is in unit of $T_r$ [Eq.~\eqref{eq:radiative_time}]. Predictions derived in this work [Eqs.~\eqref{eq:st:N1exact} and~\eqref{eq:lt:Nn:approx2}] are shown as black lines.
    (c) Pair spectra at early time $t=10^{-3}\,T_r$ as a function of $\gamma/\gamma_0$ for the three first generations. Color lines stand for MC simulations and black ones for our theoretical prediction Eq.~\eqref{eq:sh:sol:fpn}. (d) Total photon spectrum at late time $t=10^3T_r$ as a function of $\gamma_\gamma/\gamma_0$. Blue and red lines represent the total and decayed spectra from MC simulation. The black solid line is obtained by summing Eq.~\eqref{eq:lt:Fnapprox} over all generations. The black dashed line is $\chi_{l}(t)/\chi_0$ from Eq.~\eqref{eq:lt:chiL}. For panels (b-d), $\gamma_0=10^5$ and $B=B_s/100$ ($\chi_0=10^3$). (e-f) Shower multiplicity as a function $\chi_0$ at $t=10^{-3}\,T_r$ (e) and $t=10^3\,T_r$ (f) obtained (red dots) from MC simulations  and (black solid lines) predicted by Eq.~\eqref{eq:st:N1exact} (at $\chi_0\gg1$) in (e) and Eq.~\eqref{eq:lt:Nn:approx2} in (f).}
    \label{fig:subplots}
\end{figure*}

As previously stated, EMS develop into several successive generations defined as follows: a lepton of generation $n$ creates photons of generation $n$ which decay into new pairs of generation $n+1$, an example is shown in Fig.~\ref{fig:subplots}(a). The temporal evolution of the energy distribution of each generation ($n$) of electrons ($\scriptstyle{-}$), positrons ($\scriptstyle{+}$) and photons ($\gamma$) reads~\cite{pouyez2024multiplicity}:
\begin{eqnarray}
    \nonumber \partial_t f_\pm^{(n)}(\gamma,t) &=& \int_0^{\infty}\!\! d\gamma_\gamma\, w(\gamma+\gamma_\gamma,\gamma_\gamma)\,f_\pm^{(n)}(\gamma+\gamma_\gamma,t) \nonumber\\
    &-& \int_0^{\infty}\!\! d\gamma_\gamma\, w(\gamma,\gamma_\gamma)\,f_\pm^{(n)}(\gamma,t) \nonumber\\
    &+&\int_0^{\infty}\!\! d\gamma_\gamma\, \overline{w}(\gamma_\gamma,\gamma)\,f_\gamma^{(n-1)}(\gamma_\gamma,t)\,, \label{eq:fpn} \\
    \nonumber \partial_t f_\gamma^{(n)}(\gamma_\gamma,t) &=& \int_1^{\infty}\!\! d\gamma\, w(\gamma,\gamma_\gamma)\,f_-^{(n)}(\gamma,t)\\
    &+&\int_1^{\infty}\!\! d\gamma\, w(\gamma,\gamma_\gamma)\,f_+^{(n)}(\gamma,t) \nonumber\\
    &-& \Wbw(\gamma_\gamma)f_\gamma^{(n)}(\gamma_\gamma,t) \, . \label{eq:fgn}
\end{eqnarray}
\noindent where $w(\gamma,\gamma_\gamma)$ and $\overline{w}(\gamma_\gamma,\gamma)$ are the (energy) differential rates of high-energy photon emission (nICS) and pair production (nBW), respectively, within the locally constant field approximation, and $\Wcs(\gamma)=\int\!d\gamma_\gamma\,w(\gamma,\gamma_\gamma)$ and 
$\Wbw(\gamma_\gamma)=\int\!d\gamma\,\overline{w}(\gamma_\gamma,\gamma)$, with $(\gamma-1)\,mc^2$ the lepton kinetic energy and $\gamma_\gamma\,mc^2$ the photon energy. Considering an EMS developing from particles (leptons and photons) propagating perpendicularly to a (constant and homogeneous) magnetic field $B$, the rates depend on $B$ through the leptons' and photons' quantum parameters $\chi \simeq \gamma\,B/B_s$ and $\chi_\gamma = \gamma_\gamma\,B/B_s$, with $B_s = m^2c^2/(e\hbar) \sim 4.4\times 10^9~{\rm T}$ the Schwinger magnetic field limit. Details on the rates and assumptions behind Eqs.~\eqref{eq:fpn} and~\eqref{eq:fgn} are given in ~\cite{SuppMat}.

Let us now focus on an EMS developing from a seed-electron with initial kinetic energy $(\gamma_0-1) mc^2$ as depicted in Fig.~\ref{fig:subplots}(a). We can define an (initial) radiation time $T_r$ corresponding to the necessary time for the seed electron to lose a significant part of its energy as it radiates in the magnetic field (see ~\cite{akhiezer1994kinetic} and ~\cite{SuppMat} for details): 
\begin{eqnarray}\label{eq:radiative_time}
    T_r = c_1\,\frac{\tau_c}{\alpha}\,\gamma_0^{1/3}\left(\frac{B_s}{B}\right)^{2/3} = c_1 \,\frac{\tau_c}{\alpha}\gamma_0\chi_0^{-2/3} \,
\end{eqnarray}
with $\tau_c = \hbar/(mc^2)$, $\alpha=e^2/(4\pi\epsilon_0\hbar c)$ and $c_1\simeq 8.09$ (throughout this work, all constants $c_i$, with $i$ from $1$ to $8$, are integration constants whose analytical forms are explicitly given in \cite{SuppMat}).

In Fig.~\ref{fig:subplots}(b), the shower multiplicity (red) and the average energy of the seed electrons (blue) are shown as a function of the normalized time $t/T_r$. These curves are extracted from MC simulations that integrate Eqs. \eqref{eq:fpn} and \eqref{eq:fgn}. The shower multiplicity exhibits two different behaviours, associated to two different regimes of evolution for the shower. At short time scales ($t \ll T_r$) the multiplicity grows quickly [$\propto (t/T_r)^2$], but the overall number of produced pairs is small:  the charged particles are likely to emit few photons and as a first approximation their energy cooling can be neglected. This regimes holds as long as the shower multiplicity is less than or of order $1$. On longer time scales, the leptons have converted all their energy into photons and only the low energetic ones still contribute to the shower, the multiplicity shows a slow [$\propto \ln (t/T_r)$] asymptotic increase. 

%\section{Short time}
%_________________________________________________________________%

In the short-time limit ($t \ll T_r$), photon emission does not impact the overall lepton dynamics and radiation reaction, i.e. the first two terms in Eq.~\eqref{eq:fpn}, can be neglected.
The integration of Eqs.~\eqref{eq:fpn} and~\eqref{eq:fgn} gives the recursive form for the distribution functions of generation $n\geq 1$:
\begin{eqnarray}
    f_\gamma^{(n-1)}(\gamma_\gamma,t)&\simeq & \frac{2^{n-1}}{(2n-1)!}G^{(n-1)}(\gamma_\gamma) \left(\frac{t}{T_r}\right)^{2n-1},\label{eq:sh:sol:fgn} \\
    f_\pm^{(n)}(\gamma,t)&\simeq& \frac{2^{n-1}}{(2n)!}L^{(n)}(\gamma)\left(\frac{t}{T_r}\right)^{2n} ,\label{eq:sh:sol:fpn} 
\end{eqnarray}
with
\begin{eqnarray}
     G^{(n)}(\gamma_\gamma)&=& T_r\int_1^{\infty}\!\! d\gamma \, w(\gamma,\gamma_\gamma)L^{(n)}(\gamma)
    ,\label{eq:sh:Gn}
\end{eqnarray}
\begin{eqnarray}
     L^{(n)}(\gamma)&=& T_r\int_0^{\infty}\!\! d\gamma_\gamma \, \overline{w}(\gamma_\gamma,\gamma)G^{(n-1)}(\gamma_\gamma), \label{eq:sh:Ln}\\
    L^{(0)}(\gamma)&=&f_-^{(0)}(\gamma,t=0).
\end{eqnarray}
The pair spectra $f_\pm^{(n)}$ obtained from Eq. \eqref{eq:sh:sol:fpn} is plotted in Fig.~\ref{fig:subplots}(c) at $t=T_r/1000$, for $\gamma_0=10^5$ and $B=B_s/100$. The black solid lines represent the theoretical prediction Eq.~\eqref{eq:sh:sol:fpn} for the three first generations. The results of the MC simulations for the three first generations are shown in blue, red and green which are in excellent agreement with the theoretical predictions.

The number of pairs of the $(n)^{th}$ generation is obtained by integrating Eq. \eqref{eq:sh:sol:fpn} in energy. Since the contribution of the $(n)^{th}$ generation scales as $(t/T_r)^{2n}$, in the short-time limit ($t\ll T_r$) the total shower multiplicity (sum over all generations) is dominated by the first generation. 

For $N_0$ initial electrons at $(\gamma_0-1)mc^2$ the multiplicity $N_\pm/N_0$ at $t$ is given by
\begin{eqnarray}
    N_\pm(t)/N_0&=&\left(\frac{t}{T_r}\right)^2\times\frac{T_r^2}{2}\int_0^\infty\!\!d\gamma_\gamma\,\overline{W}(\gamma_\gamma)\,w(\gamma_0,\gamma_\gamma)\label{eq:st:N1exact} \\
    &\simeq& \left(\frac{t}{T_r}\right)^2\!\! \times
    \begin{cases}
        1.25\chi_0^{2/3}e^{-16/(3\chi_0)}, &\!\!{\rm for}~\chi_0 \lesssim  1\\
        6.1\ln(\chi_0)-13.6, &\!\!{\rm for}~\chi_0 \gg 1
    \end{cases}
    \nonumber
\end{eqnarray}

The solution Eq. \eqref{eq:st:N1exact} at $\chi_0\gg1$ is plotted in black solid line as a function of time (for $\chi_0=10^3$) in Fig.~\ref{fig:subplots}(b) and as function of $\chi_0$ (for $t=T_r/1000$) in Fig.~\ref{fig:subplots}(e). 
In both cases, an excellent agreement is found between MC and Eq.~\eqref{eq:st:N1exact}.  It is worth noting that, as  visible in Fig.~\ref{fig:subplots}(b), the agreement is very good up to times $t\lesssim T_r$.
In addition, we want to stress that; in Fig.~\ref{fig:subplots}(e), for each MC simulation (red point) we considered two different couples of $B$ and $\gamma_0$ for a given $\chi_0$. The same results are obtained in both cases confirming that the parameters $\chi_0$ and $t/T_r$ are sufficient to describe the problem. 

%\section{Long time}
%_________________________________________________________________%

Let us now turn to the long-time limit ($t\gg T_r$).
On such timescales, the leptons have already lost most of their energy to radiation and mainly low energetic photons remain that slowly decay in the magnetic field. The photons that decay have been created at a much earlier time $t'$ with respect to $t$ and their probability to have decayed at time $t$ is well approximated by taking $t’=0$.  The number of pairs of generation $n$ then writes ~\cite{SuppMat}:
\begin{eqnarray}\label{eq:lt:Nn}
    N_\pm^{(n)}(t)\simeq\int_0^\infty\!\!d\gamma_\gamma\,F_{\gamma}^{(n-1)}(\gamma_\gamma,t)\,P(\gamma_\gamma,t)
\end{eqnarray}
where $F_{\gamma}^{(n-1)}(\gamma_\gamma,t)$ stands for the energy distribution of all emitted photons of generation $n-1$ at time $t$, 
and $P(\gamma_\gamma,t)=1-\exp[-\overline{W}(\gamma_\gamma)t]$ denotes the probability for a photon of energy $\gamma_\gamma$ to decay into a pair over a time $t$. 

To compute $F_{\gamma}^{(n-1)}(\gamma_\gamma,t)$, we first construct the energy spectrum $I_\gamma(\gamma,\gamma_\gamma)$ of all photons emitted by a lepton with initial energy $\gamma$ from its time of creation to a (much) later time $t$ at which the lepton has lost all its energy to photons. As derived in~\cite{SuppMat}, this defines the transfer function
\begin{eqnarray}\label{eq:lt:transfert}
    I_\gamma(\gamma,\gamma_\gamma) \simeq \frac{3}{2}\frac{\tau_c}{\alpha} \, \int_{1}^{\gamma}\!d\gamma'\, \frac{w(\gamma,\gamma_\gamma)}{\chi'^2g(\chi')}
\end{eqnarray}
for $\gamma_\gamma<\gamma$ and $0$ otherwise, and where $g(\chi')$ is the Gaunt factor of quantum radiation reaction and $\chi'=\gamma' B/B_s$. 

The recursive derivation of $F_{\gamma}^{(n-1)}(\gamma_\gamma,t)$ is given in \cite{SuppMat} and can be understand as follows. The generation $(n-2)$ of photons create pairs, and each of these pairs creates the generation $(n-1)$ of photons according to the transfer function $I_\gamma(\gamma_\gamma,\gamma)$. The spectrum of pairs at the moment of their creation is obtained from Eqs.~\eqref{eq:fpn} and~\eqref{eq:fgn} removing the radiations operators and can be expressed as a function of the total photon spectrum $F_{\gamma}^{(n-2)}(\gamma_\gamma,t)$.
\begin{eqnarray}\label{eq:lt:Fn}
    F_\gamma^{(n-1)}(\gamma_\gamma,t) &\simeq&  2\int_{0}^{\infty}\!\!\!d\gamma\, I_\gamma(\gamma,\gamma_\gamma)\int_{0}^{\infty} d \gamma_\gamma' \, \frac{\overline{w}_(\gamma_\gamma',\gamma)}{\overline{W}(\gamma_\gamma')} \nonumber \\
    &\times& F_\gamma^{(n-2)}(\gamma_\gamma',t) P(\gamma_\gamma',t)\,.
\end{eqnarray}

\begin{figure*}[htbp]
    \centering
    \includegraphics[width=\textwidth]{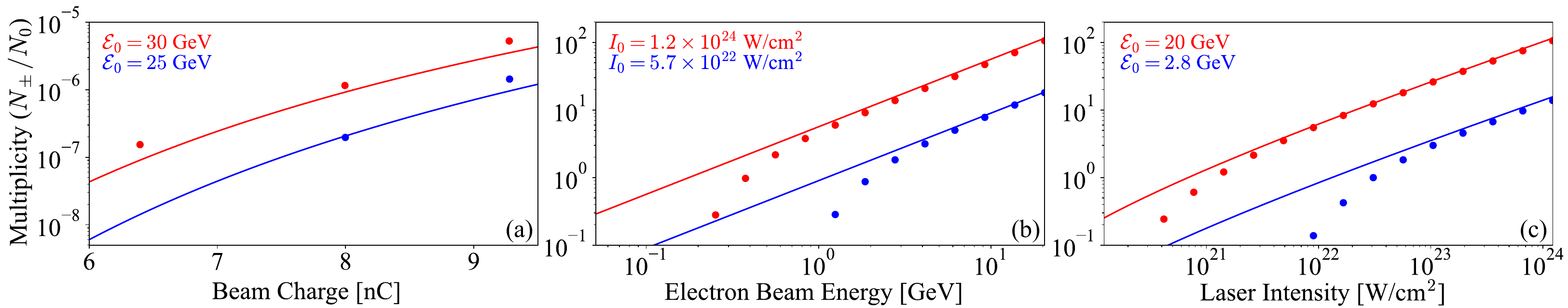} 
    \caption{Multiplicity of SF-QED showers for laboratory experiments. (a) Collision of two electron beams of energy $\mathcal{E}_0$,  as a function of the beam charge. Dots are particle-in-cell (PIC) simulation results from~\cite{del2019bright} and solid lines theoretical predictions from Eq.~\eqref{eq:beambeam:N}.  (b) and (c) Collision of an electron beam with a multi-petawatt-class laser pulse (with duration $T_{\rm\scriptscriptstyle{L}}=150~{\rm fs}$): (b) as a function of the electron beam energy for two laser intensities $I_0$, (c) as a function of the laser intensity for two electron beam energies $\mathcal{E}_0$. Dots are the results of one-dimensional PIC simulations and solid lines theoretical predictions from Eq.~\eqref{eq:lt:Nn:approx2}. 
    }
    \label{fig:subplots_appli}
\end{figure*}

This equation can be simplified noting that, in the asymptotic regime ($t\gg T_r$), mainly low energy photons remain which decay equally splitting their energy between the two resulting leptons so that $\overline{w}(\gamma_\gamma,\gamma) \simeq \overline{W}(\gamma_\gamma)\,\delta(\gamma-\gamma_\gamma/2)$.
In addition, $P(\gamma_\gamma,t)$ can be represented with high accuracy (see~\cite{SuppMat}) as an Heaviside function $\Theta(\gamma_\gamma-\gamma_{l}(t))$
with $\chi_l(t) = \gamma_{l}(t)\,B/B_s$ the photon quantum parameter below which pair production is negligible. In previous works~\cite{akhiezer1994kinetic,qu2022collective}, $\chi_l(t) \sim 1$ was intuitively considered. Here, we introduce a general, time dependent condition by defining $\chi_l(t)$ such that $\Wbw\!\left(\gamma_{l}(t)\right)t=1$. Considering $\chi_l \lesssim 1$, we obtain:
\begin{eqnarray}\label{eq:lt:chiL}
\chi_{l}(t)=\frac{8}{3\,\ln(c_2 \chi_0^{1/3}t/T_r)}\,,
\end{eqnarray}
\noindent where $c_2\simeq1.86$. It follows that Eq. \eqref{eq:lt:Fn} simplifies to:
\begin{eqnarray}\label{eq:lt:Fnapprox}
    F_\gamma^{(n-1)}(\gamma_\gamma,t) \simeq 2\!\int_{\gamma_{l}(t)}^{\gamma_0} \!\!\!d\gamma_\gamma'\, I_\gamma\!\left(\gamma_\gamma'/2,\gamma_\gamma\right)F_\gamma^{(n-2)}(\gamma_\gamma',t).\,\,
\end{eqnarray}
Equation ~\eqref{eq:lt:Fnapprox} represent accurately the 
the total spectrum of emitted photons as shown in Fig.~\ref{fig:subplots}(d). 
When injected into Eq.~\eqref{eq:lt:Nn}, it leads for the number of pairs of the $n^{th}$ generation~\cite{SuppMat}:
\begin{eqnarray}\label{eq:lt:Nn:approx1}
    &N&\!\!\!\!^{(n)}_\pm(t)/N_0\simeq c_3\, c_4^{n-1}\times \Theta\left[\chi_0-\chi_{l}(t)c_5^{n-1}\right]\quad\,\,\\
     &\times&\left[\ln\left(\frac{\chi_0}{\chi_{l}(t)}\right) -c_6(n-1) +3\,c_7^{n-1}\left(\frac{\chi_{l}(t)}{\chi_0}\right)^{1/3}-3 \right],\nonumber
\end{eqnarray}
with $c_3\simeq4.22$, $c_4\simeq 4.58$, $c_5\simeq 4.74$, $c_6\simeq 1.56$ and $c_7\simeq 1.68$. 
It follows that the generation $n$ of pairs exists only if:
\begin{eqnarray}\label{eq:number_of_gen}
n<1+c_6^{-1}\ln(\chi_0/\chi_{l}(t)).
\end{eqnarray}
This demonstrates one of the main characteristics of the shower-type cascades: 
only a finite number of generations contribute to the shower,
and this number 
$\propto \ln\big(\chi_0 \ln(\chi_0^{1/3}t/T_r)\big)$ evolves slowly with the governing parameters.
This differs from avalanche-type cascades~\cite{mironov2014collapse,mercuri2024growth} in which leptons can be constantly accelerated and where the number of generations should at minima evolve linearly with time. 

Finally, the shower multiplicity is obtained summing Eq.~\eqref{eq:number_of_gen} over all contributing generations leading for $\chi_0/\chi_{l}(t) \gg 1$~\cite{SuppMat}:
\begin{eqnarray}\label{eq:lt:Nn:approx2}
    N_\pm(t)/N_0\simeq c_8\,\chi_0\ln{\left(c_2 \chi_0^{1/3}\frac{t}{T_r}\right)}.
\end{eqnarray}
with $c_8\simeq 0.26$. 
This result can be rewritten as $N_\pm(t) \propto \chi_0/\chi_{l}(t)\propto \gamma_0/\gamma_l(t)$. The slow logarithmic time dependence is related to $\chi_{l}(t)$ while the leading dependency $\propto \chi_0$, is consistent with previous findings~\cite{heitler1984quantum,akhiezer1994kinetic,qu2022collective,selivanov2024final} and easily recovered considering $\chi_l \sim 1$.

The long-time predictions are tested against MC simulations in Figs.~\ref{fig:subplots}(b), (d) and~(f). In panel (d), the total photon spectrum emitted up to $t=10^{3}\,T_r$ is shown for $B=B_s/100$ and $\gamma_0=10^5$. The spectrum extracted from a MC simulation (in blue) is found to be in excellent agreement with the prediction of Eq.~\eqref{eq:lt:Fnapprox} (summed over all generations, in black).
The spectrum of all decayed photons, in red, is found to lay beyond the energy $\gamma_l(t)$ (vertical black dashed line) predicted using Eq.~\eqref{eq:lt:chiL}, as expected from our derivation. As a result, our solutions accurately describe the shower multiplicity at $t\gg T_r$, as shown in panel (b), which can not be computed from the work of \cite{akhiezer1994kinetic}. Finally, in panel~(f), the shower multiplicity is shown as a function of $\chi_0$. Again, excellent agreement is obtained between results from MC simulations (red dots) and the theoretical prediction of Eq.~\eqref{eq:lt:Nn:approx2}, but for a slight overestimate, also visible in panel~(b), which follows from the saddle point approximation used to derive Eq.~\eqref{eq:lt:Nn:approx1}.

%\section{Application}
%_________________________________________________________________%

The solutions derived in this work apply to any crossed EM fields for which particle re-acceleration is negligible and where the (possibly time-dependent) EM field can be associated to an equivalent constant magnetic field~\cite{di2018implementing,artemenkoPPCF2019}. 
Here we apply our results to two laboratory settings that have recently attracted much interest.

First, we discuss the collision of two uniformly charged, cylindrical electron beams of density $n_0$, radius $R$, length $L$ and energy $\mathcal{E}_0$. As the electric field in this cylinder increases linearly with the distance $r$ from its center, both the quantum parameter $\chi(r)=\chi_0\,r/R$ and the radiation time $T_r(r)=T_r(R) (R/r)^{2/3}$ are functions of $r$, and we have introduced $\chi_0 \simeq 26.8 \times \left[\mathcal{E}_0/(10~{\rm GeV})\right]\,\left[n_0/(10^{24} {\rm cm}^{-3})\right]\,
\left[R/(0.1 {\rm \mu m})\right]$ and $T_r(R) \simeq 3.12~{\rm fs} \times \left[\mathcal{E}_0/(10 \, \rm{GeV})\right]^{1/3}\left[n_0/(10^{24}\,\rm{cm}^{-3})\right]^{-2/3}$ $\left[R/(0.1 \, \mu\rm{m})\right]^{-2/3}$. 
Under conditions relevant to currently envisioned experiments~\cite{del2019bright}, one finds that $\chi_0 \lesssim 1$ and $ L/(2c)< T_r(R)$ [with $L/(2c)$ the maximum interaction time]. Hence, the resulting multiplicity can be obtained by integrating in space the short-time prediction Eq.~\eqref{eq:st:N1exact}, computed in the small-$\chi_0$ limit leading:
\begin{eqnarray}\label{eq:beambeam:N}
    N_\pm/N_0\simeq c_9\,\left(\frac{L}{c\,T_r(R)}\right)^2\,\chi_0^{-10/3}\,\Gamma\left(-4,\frac{16}{3\chi_0}\right)\,,
\end{eqnarray}
with $c_9 \simeq 1.01\cdot10^{3}$ and $\Gamma$ the upper incomplete gamma function. This prediction is compared to the numerical results from~\cite{del2019bright} in Fig.~\ref{fig:subplots_appli}(a), using the equivalent cylinder approximation~\cite{zhang2023signatures}, showing excellent agreement.

Finally, we consider the head-on collision of a beam of electrons of energy $\mathcal{E}_0$ and a laser pulse with peak intensity $I_0$ (peak electric field $E_0$), duration $T_{\rm\scriptscriptstyle{L}}$ (full-width at half-maximum in intensity) and $\sin^2$ time envelope (in intensity). Further considering $T_{\rm\scriptscriptstyle{L}} \gg \lambda/c$ (with $\lambda$ the laser wavelength), the characteristic quantum parameter and radiation time can be computed at the average electric field $\langle E \rangle=4 E_0/\pi^2$, leading $\chi_0 \simeq 5.17 \left[ \mathcal{E}_0/(10\,  \rm{ GeV})\right][I_0/(10^{22} \, \rm{ W/cm}^2)]^{1/2}$ and $T_r\simeq 9.35 \, \rm{ fs} \times \left[ \mathcal{E}_0/(10 \rm{ GeV})\right]^{1/3}[I_0/(10^{22} \, \rm{ W/cm}^2)]^{-1/3}$. Considering pulses as the ones delivered by the L4 ATON laser at ELI Beamlines, we have $T_{\rm\scriptscriptstyle{L}}\simeq 150 {\rm fs} \gg T_r$ for a wide range of parameters so that one can compute the final shower multiplicity using the long-time-limit Eq.~\eqref{eq:lt:Nn:approx2} at time $t=T_{\rm\scriptscriptstyle{L}}$. A comparison with particle-in-cell~\cite{smilei_derouillat_2018} simulations as a function of $\mathcal{E}_0$ in Fig.~\ref{fig:subplots_appli}(b) and $I_0$ in Fig.~\ref{fig:subplots_appli}(c). An excellent agreement is found for
$\chi_{l}(T_{\rm\scriptscriptstyle{L}}) \ll \chi_0$ as expected from the domain of validity of Eq. \eqref{eq:lt:Nn:approx2}.

%\section{Conclusion}
%_________________________________________________________________%

In conclusion, a complete, kinetic description of electron-seeded SF-QED showers in crossed fields is derived. It provides us with the time-dependent photon spectrum and compact analytical solutions for the shower multiplicity in short and long-time regimes. Our results are in remarkable agreement with MC simulations, while providing an explicit answer for short time, overlooked before. This work allows to derive simple scaling laws for SF-QED showers in various environments, as illustrated here by two laboratory examples.

\section*{Acknowledgements}

The authors acknowledge useful discussions with A.A. Mironov and S. Meuren.
T.G. was supported by FCT (Portugal) Grants No. CEECIND/04050/2021 and PTDC/FISPLA/ 3800/2021


\begin{thebibliography}{31}%
\makeatletter
\providecommand \@ifxundefined [1]{%
 \@ifx{#1\undefined}
}%
\providecommand \@ifnum [1]{%
 \ifnum #1\expandafter \@firstoftwo
 \else \expandafter \@secondoftwo
 \fi
}%
\providecommand \@ifx [1]{%
 \ifx #1\expandafter \@firstoftwo
 \else \expandafter \@secondoftwo
 \fi
}%
\providecommand \natexlab [1]{#1}%
\providecommand \enquote  [1]{``#1''}%
\providecommand \bibnamefont  [1]{#1}%
\providecommand \bibfnamefont [1]{#1}%
\providecommand \citenamefont [1]{#1}%
\providecommand \href@noop [0]{\@secondoftwo}%
\providecommand \href [0]{\begingroup \@sanitize@url \@href}%
\providecommand \@href[1]{\@@startlink{#1}\@@href}%
\providecommand \@@href[1]{\endgroup#1\@@endlink}%
\providecommand \@sanitize@url [0]{\catcode `\\12\catcode `\$12\catcode `\&12\catcode `\#12\catcode `\^12\catcode `\_12\catcode `\%12\relax}%
\providecommand \@@startlink[1]{}%
\providecommand \@@endlink[0]{}%
\providecommand \url  [0]{\begingroup\@sanitize@url \@url }%
\providecommand \@url [1]{\endgroup\@href {#1}{\urlprefix }}%
\providecommand \urlprefix  [0]{URL }%
\providecommand \Eprint [0]{\href }%
\providecommand \doibase [0]{https://doi.org/}%
\providecommand \selectlanguage [0]{\@gobble}%
\providecommand \bibinfo  [0]{\@secondoftwo}%
\providecommand \bibfield  [0]{\@secondoftwo}%
\providecommand \translation [1]{[#1]}%
\providecommand \BibitemOpen [0]{}%
\providecommand \bibitemStop [0]{}%
\providecommand \bibitemNoStop [0]{.\EOS\space}%
\providecommand \EOS [0]{\spacefactor3000\relax}%
\providecommand \BibitemShut  [1]{\csname bibitem#1\endcsname}%
\let\auto@bib@innerbib\@empty
%</preamble>
\bibitem [{\citenamefont {Bhabha}\ and\ \citenamefont {Heitler}(1937)}]{bhabha1937passage}%
  \BibitemOpen
  \bibfield  {author} {\bibinfo {author} {\bibfnamefont {H.~J.}\ \bibnamefont {Bhabha}}\ and\ \bibinfo {author} {\bibfnamefont {W.}~\bibnamefont {Heitler}},\ }\bibfield  {title} {\bibinfo {title} {The passage of fast electrons and the theory of cosmic showers},\ }\href@noop {} {\bibfield  {journal} {\bibinfo  {journal} {Proceedings of the Royal Society of London. Series A-Mathematical and Physical Sciences}\ }\textbf {\bibinfo {volume} {159}},\ \bibinfo {pages} {432} (\bibinfo {year} {1937})}\BibitemShut {NoStop}%
\bibitem [{\citenamefont {Carlson}\ and\ \citenamefont {Oppenheimer}(1937)}]{carlson1937multiplicative}%
  \BibitemOpen
  \bibfield  {author} {\bibinfo {author} {\bibfnamefont {J.}~\bibnamefont {Carlson}}\ and\ \bibinfo {author} {\bibfnamefont {J.}~\bibnamefont {Oppenheimer}},\ }\bibfield  {title} {\bibinfo {title} {On multiplicative showers},\ }\href@noop {} {\bibfield  {journal} {\bibinfo  {journal} {Physical Review}\ }\textbf {\bibinfo {volume} {51}},\ \bibinfo {pages} {220} (\bibinfo {year} {1937})}\BibitemShut {NoStop}%
\bibitem [{\citenamefont {Landau}\ and\ \citenamefont {Rumer}(1938)}]{landau1938cascade}%
  \BibitemOpen
  \bibfield  {author} {\bibinfo {author} {\bibfnamefont {L.~D.}\ \bibnamefont {Landau}}\ and\ \bibinfo {author} {\bibfnamefont {G.}~\bibnamefont {Rumer}},\ }\bibfield  {title} {\bibinfo {title} {The cascade theory of electronic showers},\ }\href@noop {} {\bibfield  {journal} {\bibinfo  {journal} {Proceedings of the Royal Society of London. Series A. Mathematical and Physical Sciences}\ }\textbf {\bibinfo {volume} {166}},\ \bibinfo {pages} {213} (\bibinfo {year} {1938})}\BibitemShut {NoStop}%
\bibitem [{\citenamefont {Chen}\ and\ \citenamefont {Fiuza}(2023)}]{HChen2023}%
  \BibitemOpen
  \bibfield  {author} {\bibinfo {author} {\bibfnamefont {H.}~\bibnamefont {Chen}}\ and\ \bibinfo {author} {\bibfnamefont {F.}~\bibnamefont {Fiuza}},\ }\bibfield  {title} {\bibinfo {title} {{Perspectives on relativistic electron–positron pair plasma experiments of astrophysical relevance using high-power lasers}},\ }\href {https://doi.org/10.1063/5.0134819} {\bibfield  {journal} {\bibinfo  {journal} {Physics of Plasmas}\ }\textbf {\bibinfo {volume} {30}},\ \bibinfo {pages} {020601} (\bibinfo {year} {2023})}\BibitemShut {NoStop}%
\bibitem [{\citenamefont {Sarri}\ \emph {et~al.}(2015)\citenamefont {Sarri}, \citenamefont {Poder}, \citenamefont {Cole}, \citenamefont {Schumaker}, \citenamefont {Piazza}, \citenamefont {Reville}, \citenamefont {Dzelzainis}, \citenamefont {Doria}, \citenamefont {Gizzi}, \citenamefont {Grittani}, \citenamefont {Kar}, \citenamefont {Keitel}, \citenamefont {Krushelnick}, \citenamefont {Kuschel}, \citenamefont {Mangles}, \citenamefont {Najmudin}, \citenamefont {Shukla}, \citenamefont {Silva}, \citenamefont {Symes}, \citenamefont {Thomas}, \citenamefont {Vargas}, \citenamefont {Vieira},\ and\ \citenamefont {Zepf}}]{generation_sarri_2015}%
  \BibitemOpen
  \bibfield  {author} {\bibinfo {author} {\bibfnamefont {G.}~\bibnamefont {Sarri}}, \bibinfo {author} {\bibfnamefont {K.}~\bibnamefont {Poder}}, \bibinfo {author} {\bibfnamefont {J.}~\bibnamefont {Cole}}, \bibinfo {author} {\bibfnamefont {W.}~\bibnamefont {Schumaker}}, \bibinfo {author} {\bibfnamefont {A.~D.}\ \bibnamefont {Piazza}}, \bibinfo {author} {\bibfnamefont {B.}~\bibnamefont {Reville}}, \bibinfo {author} {\bibfnamefont {T.}~\bibnamefont {Dzelzainis}}, \bibinfo {author} {\bibfnamefont {D.}~\bibnamefont {Doria}}, \bibinfo {author} {\bibfnamefont {L.~A.}\ \bibnamefont {Gizzi}}, \bibinfo {author} {\bibfnamefont {G.}~\bibnamefont {Grittani}}, \bibinfo {author} {\bibfnamefont {S.}~\bibnamefont {Kar}}, \bibinfo {author} {\bibfnamefont {C.~H.}\ \bibnamefont {Keitel}}, \bibinfo {author} {\bibfnamefont {K.}~\bibnamefont {Krushelnick}}, \bibinfo {author} {\bibfnamefont {S.}~\bibnamefont {Kuschel}}, \bibinfo {author} {\bibfnamefont {S.}~\bibnamefont {Mangles}}, \bibinfo {author} {\bibfnamefont {Z.}~\bibnamefont
  {Najmudin}}, \bibinfo {author} {\bibfnamefont {N.}~\bibnamefont {Shukla}}, \bibinfo {author} {\bibfnamefont {L.~O.}\ \bibnamefont {Silva}}, \bibinfo {author} {\bibfnamefont {D.}~\bibnamefont {Symes}}, \bibinfo {author} {\bibfnamefont {A.}~\bibnamefont {Thomas}}, \bibinfo {author} {\bibfnamefont {M.}~\bibnamefont {Vargas}}, \bibinfo {author} {\bibfnamefont {J.}~\bibnamefont {Vieira}},\ and\ \bibinfo {author} {\bibfnamefont {M.}~\bibnamefont {Zepf}},\ }\bibfield  {title} {\bibinfo {title} {Generation of neutral and high-density electron-positron pair plasmas in the laboratory},\ }\bibfield  {journal} {\bibinfo  {journal} {Nature Communications}\ }\href {https://doi.org/10.1038/NCOMMS7747} {10.1038/NCOMMS7747} (\bibinfo {year} {2015})\BibitemShut {NoStop}%
\bibitem [{\citenamefont {Arrowsmith}\ \emph {et~al.}(2024)\citenamefont {Arrowsmith}, \citenamefont {Simon}, \citenamefont {Bilbao}, \citenamefont {Bott}, \citenamefont {Burger}, \citenamefont {Chen}, \citenamefont {Cruz}, \citenamefont {Davenne}, \citenamefont {Efthymiopoulos}, \citenamefont {Froula}, \citenamefont {Goillot}, \citenamefont {Gudmundsson}, \citenamefont {Haberberger}, \citenamefont {Halliday}, \citenamefont {Hodge}, \citenamefont {Huffman}, \citenamefont {Iaquinta}, \citenamefont {Miniati}, \citenamefont {Reville}, \citenamefont {Sarkar}, \citenamefont {Schekochihin}, \citenamefont {Silva}, \citenamefont {Simpson}, \citenamefont {Stergiou}, \citenamefont {Trines}, \citenamefont {Vieu}, \citenamefont {Charitonidis}, \citenamefont {Bingham},\ and\ \citenamefont {Gregori}}]{Arrowsmith_2024}%
  \BibitemOpen
  \bibfield  {author} {\bibinfo {author} {\bibfnamefont {C.~D.}\ \bibnamefont {Arrowsmith}}, \bibinfo {author} {\bibfnamefont {P.}~\bibnamefont {Simon}}, \bibinfo {author} {\bibfnamefont {P.~J.}\ \bibnamefont {Bilbao}}, \bibinfo {author} {\bibfnamefont {A.~F.~A.}\ \bibnamefont {Bott}}, \bibinfo {author} {\bibfnamefont {S.}~\bibnamefont {Burger}}, \bibinfo {author} {\bibfnamefont {H.}~\bibnamefont {Chen}}, \bibinfo {author} {\bibfnamefont {F.~D.}\ \bibnamefont {Cruz}}, \bibinfo {author} {\bibfnamefont {T.}~\bibnamefont {Davenne}}, \bibinfo {author} {\bibfnamefont {I.}~\bibnamefont {Efthymiopoulos}}, \bibinfo {author} {\bibfnamefont {D.~H.}\ \bibnamefont {Froula}}, \bibinfo {author} {\bibfnamefont {A.}~\bibnamefont {Goillot}}, \bibinfo {author} {\bibfnamefont {J.~T.}\ \bibnamefont {Gudmundsson}}, \bibinfo {author} {\bibfnamefont {D.}~\bibnamefont {Haberberger}}, \bibinfo {author} {\bibfnamefont {J.~W.~D.}\ \bibnamefont {Halliday}}, \bibinfo {author} {\bibfnamefont {T.}~\bibnamefont {Hodge}}, \bibinfo {author}
  {\bibfnamefont {B.~T.}\ \bibnamefont {Huffman}}, \bibinfo {author} {\bibfnamefont {S.}~\bibnamefont {Iaquinta}}, \bibinfo {author} {\bibfnamefont {F.}~\bibnamefont {Miniati}}, \bibinfo {author} {\bibfnamefont {B.}~\bibnamefont {Reville}}, \bibinfo {author} {\bibfnamefont {S.}~\bibnamefont {Sarkar}}, \bibinfo {author} {\bibfnamefont {A.~A.}\ \bibnamefont {Schekochihin}}, \bibinfo {author} {\bibfnamefont {L.~O.}\ \bibnamefont {Silva}}, \bibinfo {author} {\bibfnamefont {R.}~\bibnamefont {Simpson}}, \bibinfo {author} {\bibfnamefont {V.}~\bibnamefont {Stergiou}}, \bibinfo {author} {\bibfnamefont {R.~M. G.~M.}\ \bibnamefont {Trines}}, \bibinfo {author} {\bibfnamefont {T.}~\bibnamefont {Vieu}}, \bibinfo {author} {\bibfnamefont {N.}~\bibnamefont {Charitonidis}}, \bibinfo {author} {\bibfnamefont {R.}~\bibnamefont {Bingham}},\ and\ \bibinfo {author} {\bibfnamefont {G.}~\bibnamefont {Gregori}},\ }\bibfield  {title} {\bibinfo {title} {Laboratory realization of relativistic pair-plasma beams},\ }\bibfield  {journal}
  {\bibinfo  {journal} {Nature Communications}\ }\textbf {\bibinfo {volume} {15}},\ \href {https://doi.org/10.1038/s41467-024-49346-2} {10.1038/s41467-024-49346-2} (\bibinfo {year} {2024})\BibitemShut {NoStop}%
\bibitem [{\citenamefont {Di~Piazza}\ \emph {et~al.}(2012)\citenamefont {Di~Piazza}, \citenamefont {M{\"u}ller}, \citenamefont {Hatsagortsyan},\ and\ \citenamefont {Keitel}}]{di2012extremely}%
  \BibitemOpen
  \bibfield  {author} {\bibinfo {author} {\bibfnamefont {A.}~\bibnamefont {Di~Piazza}}, \bibinfo {author} {\bibfnamefont {C.}~\bibnamefont {M{\"u}ller}}, \bibinfo {author} {\bibfnamefont {K.}~\bibnamefont {Hatsagortsyan}},\ and\ \bibinfo {author} {\bibfnamefont {C.~H.}\ \bibnamefont {Keitel}},\ }\bibfield  {title} {\bibinfo {title} {Extremely high-intensity laser interactions with fundamental quantum systems},\ }\href@noop {} {\bibfield  {journal} {\bibinfo  {journal} {Reviews of Modern Physics}\ }\textbf {\bibinfo {volume} {84}},\ \bibinfo {pages} {1177} (\bibinfo {year} {2012})}\BibitemShut {NoStop}%
\bibitem [{\citenamefont {Gonoskov}\ \emph {et~al.}(2022)\citenamefont {Gonoskov}, \citenamefont {Blackburn}, \citenamefont {Marklund},\ and\ \citenamefont {Bulanov}}]{gonoskov2022charged}%
  \BibitemOpen
  \bibfield  {author} {\bibinfo {author} {\bibfnamefont {A.}~\bibnamefont {Gonoskov}}, \bibinfo {author} {\bibfnamefont {T.}~\bibnamefont {Blackburn}}, \bibinfo {author} {\bibfnamefont {M.}~\bibnamefont {Marklund}},\ and\ \bibinfo {author} {\bibfnamefont {S.}~\bibnamefont {Bulanov}},\ }\bibfield  {title} {\bibinfo {title} {Charged particle motion and radiation in strong electromagnetic fields},\ }\href@noop {} {\bibfield  {journal} {\bibinfo  {journal} {Reviews of Modern Physics}\ }\textbf {\bibinfo {volume} {94}},\ \bibinfo {pages} {045001} (\bibinfo {year} {2022})}\BibitemShut {NoStop}%
\bibitem [{\citenamefont {Fedotov}\ \emph {et~al.}(2023)\citenamefont {Fedotov}, \citenamefont {Ilderton}, \citenamefont {Karbstein}, \citenamefont {King}, \citenamefont {Seipt}, \citenamefont {Taya},\ and\ \citenamefont {Torgrimsson}}]{fedotov2023advances}%
  \BibitemOpen
  \bibfield  {author} {\bibinfo {author} {\bibfnamefont {A.}~\bibnamefont {Fedotov}}, \bibinfo {author} {\bibfnamefont {A.}~\bibnamefont {Ilderton}}, \bibinfo {author} {\bibfnamefont {F.}~\bibnamefont {Karbstein}}, \bibinfo {author} {\bibfnamefont {B.}~\bibnamefont {King}}, \bibinfo {author} {\bibfnamefont {D.}~\bibnamefont {Seipt}}, \bibinfo {author} {\bibfnamefont {H.}~\bibnamefont {Taya}},\ and\ \bibinfo {author} {\bibfnamefont {G.}~\bibnamefont {Torgrimsson}},\ }\bibfield  {title} {\bibinfo {title} {Advances in qed with intense background fields},\ }\href@noop {} {\bibfield  {journal} {\bibinfo  {journal} {Physics Reports}\ }\textbf {\bibinfo {volume} {1010}},\ \bibinfo {pages} {1} (\bibinfo {year} {2023})}\BibitemShut {NoStop}%
\bibitem [{\citenamefont {Sturrock}(1971)}]{sturrock1971model}%
  \BibitemOpen
  \bibfield  {author} {\bibinfo {author} {\bibfnamefont {P.}~\bibnamefont {Sturrock}},\ }\bibfield  {title} {\bibinfo {title} {A model of pulsars},\ }\href@noop {} {\bibfield  {journal} {\bibinfo  {journal} {Astrophysical Journal, vol. 164, p. 529}\ }\textbf {\bibinfo {volume} {164}},\ \bibinfo {pages} {529} (\bibinfo {year} {1971})}\BibitemShut {NoStop}%
\bibitem [{\citenamefont {Ruderman}\ and\ \citenamefont {Sutherland}(1975)}]{ruderman1975theory}%
  \BibitemOpen
  \bibfield  {author} {\bibinfo {author} {\bibfnamefont {M.}~\bibnamefont {Ruderman}}\ and\ \bibinfo {author} {\bibfnamefont {P.~G.}\ \bibnamefont {Sutherland}},\ }\bibfield  {title} {\bibinfo {title} {Theory of pulsars-polar caps, sparks, and coherent microwave radiation},\ }\href@noop {} {\bibfield  {journal} {\bibinfo  {journal} {Astrophysical Journal, vol. 196, Feb. 15, 1975, pt. 1, p. 51-72.}\ }\textbf {\bibinfo {volume} {196}},\ \bibinfo {pages} {51} (\bibinfo {year} {1975})}\BibitemShut {NoStop}%
\bibitem [{\citenamefont {Akhiezer}\ \emph {et~al.}(1994)\citenamefont {Akhiezer}, \citenamefont {Merenkov},\ and\ \citenamefont {Rekalo}}]{akhiezer1994kinetic}%
  \BibitemOpen
  \bibfield  {author} {\bibinfo {author} {\bibfnamefont {A.}~\bibnamefont {Akhiezer}}, \bibinfo {author} {\bibfnamefont {N.}~\bibnamefont {Merenkov}},\ and\ \bibinfo {author} {\bibfnamefont {A.}~\bibnamefont {Rekalo}},\ }\bibfield  {title} {\bibinfo {title} {On a kinetic theory of electromagnetic showers in strong magnetic fields},\ }\href@noop {} {\bibfield  {journal} {\bibinfo  {journal} {Journal of Physics G: Nuclear and Particle Physics}\ }\textbf {\bibinfo {volume} {20}},\ \bibinfo {pages} {1499} (\bibinfo {year} {1994})}\BibitemShut {NoStop}%
\bibitem [{\citenamefont {Hibschman}\ and\ \citenamefont {Arons}(2001{\natexlab{a}})}]{hibschman2001pair1}%
  \BibitemOpen
  \bibfield  {author} {\bibinfo {author} {\bibfnamefont {J.~A.}\ \bibnamefont {Hibschman}}\ and\ \bibinfo {author} {\bibfnamefont {J.}~\bibnamefont {Arons}},\ }\bibfield  {title} {\bibinfo {title} {Pair multiplicities and pulsar death},\ }\href@noop {} {\bibfield  {journal} {\bibinfo  {journal} {The Astrophysical Journal}\ }\textbf {\bibinfo {volume} {554}},\ \bibinfo {pages} {624} (\bibinfo {year} {2001}{\natexlab{a}})}\BibitemShut {NoStop}%
\bibitem [{\citenamefont {Hibschman}\ and\ \citenamefont {Arons}(2001{\natexlab{b}})}]{hibschman2001pair2}%
  \BibitemOpen
  \bibfield  {author} {\bibinfo {author} {\bibfnamefont {J.~A.}\ \bibnamefont {Hibschman}}\ and\ \bibinfo {author} {\bibfnamefont {J.}~\bibnamefont {Arons}},\ }\bibfield  {title} {\bibinfo {title} {Pair production multiplicities in rotation-powered pulsars},\ }\href@noop {} {\bibfield  {journal} {\bibinfo  {journal} {The Astrophysical Journal}\ }\textbf {\bibinfo {volume} {560}},\ \bibinfo {pages} {871} (\bibinfo {year} {2001}{\natexlab{b}})}\BibitemShut {NoStop}%
\bibitem [{\citenamefont {Anguelov}\ and\ \citenamefont {Vankov}(1999)}]{anguelov1999electromagnetic}%
  \BibitemOpen
  \bibfield  {author} {\bibinfo {author} {\bibfnamefont {V.}~\bibnamefont {Anguelov}}\ and\ \bibinfo {author} {\bibfnamefont {H.}~\bibnamefont {Vankov}},\ }\bibfield  {title} {\bibinfo {title} {Electromagnetic showers in a strong magnetic field},\ }\href@noop {} {\bibfield  {journal} {\bibinfo  {journal} {Journal of Physics G: Nuclear and Particle Physics}\ }\textbf {\bibinfo {volume} {25}},\ \bibinfo {pages} {1755} (\bibinfo {year} {1999})}\BibitemShut {NoStop}%
\bibitem [{\citenamefont {Blackburn}\ \emph {et~al.}(2017)\citenamefont {Blackburn}, \citenamefont {Ilderton}, \citenamefont {Murphy},\ and\ \citenamefont {Marklund}}]{blackburn2017scaling}%
  \BibitemOpen
  \bibfield  {author} {\bibinfo {author} {\bibfnamefont {T.}~\bibnamefont {Blackburn}}, \bibinfo {author} {\bibfnamefont {A.}~\bibnamefont {Ilderton}}, \bibinfo {author} {\bibfnamefont {C.}~\bibnamefont {Murphy}},\ and\ \bibinfo {author} {\bibfnamefont {M.}~\bibnamefont {Marklund}},\ }\bibfield  {title} {\bibinfo {title} {Scaling laws for positron production in laser--electron-beam collisions},\ }\href@noop {} {\bibfield  {journal} {\bibinfo  {journal} {Physical Review A}\ }\textbf {\bibinfo {volume} {96}},\ \bibinfo {pages} {022128} (\bibinfo {year} {2017})}\BibitemShut {NoStop}%
\bibitem [{\citenamefont {Mercuri-Baron}\ \emph {et~al.}(2021)\citenamefont {Mercuri-Baron}, \citenamefont {Grech}, \citenamefont {Niel}, \citenamefont {Grassi}, \citenamefont {Lobet}, \citenamefont {Di~Piazza},\ and\ \citenamefont {Riconda}}]{mercuri2021impact}%
  \BibitemOpen
  \bibfield  {author} {\bibinfo {author} {\bibfnamefont {A.}~\bibnamefont {Mercuri-Baron}}, \bibinfo {author} {\bibfnamefont {M.}~\bibnamefont {Grech}}, \bibinfo {author} {\bibfnamefont {F.}~\bibnamefont {Niel}}, \bibinfo {author} {\bibfnamefont {A.}~\bibnamefont {Grassi}}, \bibinfo {author} {\bibfnamefont {M.}~\bibnamefont {Lobet}}, \bibinfo {author} {\bibfnamefont {A.}~\bibnamefont {Di~Piazza}},\ and\ \bibinfo {author} {\bibfnamefont {C.}~\bibnamefont {Riconda}},\ }\bibfield  {title} {\bibinfo {title} {Impact of the laser spatio-temporal shape on breit--wheeler pair production},\ }\href@noop {} {\bibfield  {journal} {\bibinfo  {journal} {New Journal of Physics}\ }\textbf {\bibinfo {volume} {23}},\ \bibinfo {pages} {085006} (\bibinfo {year} {2021})}\BibitemShut {NoStop}%
\bibitem [{\citenamefont {Lobet}\ \emph {et~al.}(2017)\citenamefont {Lobet}, \citenamefont {Davoine}, \citenamefont {d’Humi{\`e}res},\ and\ \citenamefont {Gremillet}}]{lobet2017generation}%
  \BibitemOpen
  \bibfield  {author} {\bibinfo {author} {\bibfnamefont {M.}~\bibnamefont {Lobet}}, \bibinfo {author} {\bibfnamefont {X.}~\bibnamefont {Davoine}}, \bibinfo {author} {\bibfnamefont {E.}~\bibnamefont {d’Humi{\`e}res}},\ and\ \bibinfo {author} {\bibfnamefont {L.}~\bibnamefont {Gremillet}},\ }\bibfield  {title} {\bibinfo {title} {Generation of high-energy electron-positron pairs in the collision of a laser-accelerated electron beam with a multipetawatt laser},\ }\href@noop {} {\bibfield  {journal} {\bibinfo  {journal} {Physical Review Accelerators and Beams}\ }\textbf {\bibinfo {volume} {20}},\ \bibinfo {pages} {043401} (\bibinfo {year} {2017})}\BibitemShut {NoStop}%
\bibitem [{\citenamefont {Qu}\ \emph {et~al.}(2022)\citenamefont {Qu}, \citenamefont {Meuren},\ and\ \citenamefont {Fisch}}]{qu2022collective}%
  \BibitemOpen
  \bibfield  {author} {\bibinfo {author} {\bibfnamefont {K.}~\bibnamefont {Qu}}, \bibinfo {author} {\bibfnamefont {S.}~\bibnamefont {Meuren}},\ and\ \bibinfo {author} {\bibfnamefont {N.~J.}\ \bibnamefont {Fisch}},\ }\bibfield  {title} {\bibinfo {title} {Collective plasma effects of electron--positron pairs in beam-driven qed cascades},\ }\href@noop {} {\bibfield  {journal} {\bibinfo  {journal} {Physics of Plasmas}\ }\textbf {\bibinfo {volume} {29}},\ \bibinfo {pages} {042117} (\bibinfo {year} {2022})}\BibitemShut {NoStop}%
\bibitem [{\citenamefont {Pouyez}\ \emph {et~al.}(2024{\natexlab{a}})\citenamefont {Pouyez}, \citenamefont {Mironov}, \citenamefont {Grismayer}, \citenamefont {Mercuri-Baron}, \citenamefont {Perez}, \citenamefont {Vranic}, \citenamefont {Riconda},\ and\ \citenamefont {Grech}}]{pouyez2024multiplicity}%
  \BibitemOpen
  \bibfield  {author} {\bibinfo {author} {\bibfnamefont {M.}~\bibnamefont {Pouyez}}, \bibinfo {author} {\bibfnamefont {A.}~\bibnamefont {Mironov}}, \bibinfo {author} {\bibfnamefont {T.}~\bibnamefont {Grismayer}}, \bibinfo {author} {\bibfnamefont {A.}~\bibnamefont {Mercuri-Baron}}, \bibinfo {author} {\bibfnamefont {F.}~\bibnamefont {Perez}}, \bibinfo {author} {\bibfnamefont {M.}~\bibnamefont {Vranic}}, \bibinfo {author} {\bibfnamefont {C.}~\bibnamefont {Riconda}},\ and\ \bibinfo {author} {\bibfnamefont {M.}~\bibnamefont {Grech}},\ }\bibfield  {title} {\bibinfo {title} {Multiplicity of electron-and photon-seeded electromagnetic showers at multi-petawatt laser facilities},\ }\href@noop {} {\bibfield  {journal} {\bibinfo  {journal} {arXiv preprint arXiv:2402.04501}\ } (\bibinfo {year} {2024}{\natexlab{a}})}\BibitemShut {NoStop}%
\bibitem [{\citenamefont {Matheron}\ \emph {et~al.}(2024)\citenamefont {Matheron}, \citenamefont {Andriyash}, \citenamefont {Davoine}, \citenamefont {Gremillet}, \citenamefont {Pouyez}, \citenamefont {Grech}, \citenamefont {Lancia}, \citenamefont {Phuoc},\ and\ \citenamefont {Corde}}]{matheron2024self}%
  \BibitemOpen
  \bibfield  {author} {\bibinfo {author} {\bibfnamefont {A.}~\bibnamefont {Matheron}}, \bibinfo {author} {\bibfnamefont {I.}~\bibnamefont {Andriyash}}, \bibinfo {author} {\bibfnamefont {X.}~\bibnamefont {Davoine}}, \bibinfo {author} {\bibfnamefont {L.}~\bibnamefont {Gremillet}}, \bibinfo {author} {\bibfnamefont {M.}~\bibnamefont {Pouyez}}, \bibinfo {author} {\bibfnamefont {M.}~\bibnamefont {Grech}}, \bibinfo {author} {\bibfnamefont {L.}~\bibnamefont {Lancia}}, \bibinfo {author} {\bibfnamefont {K.~T.}\ \bibnamefont {Phuoc}},\ and\ \bibinfo {author} {\bibfnamefont {S.}~\bibnamefont {Corde}},\ }\bibfield  {title} {\bibinfo {title} {Self-triggered strong-field qed collisions in laser-plasma interaction},\ }\href@noop {} {\bibfield  {journal} {\bibinfo  {journal} {arXiv preprint arXiv:2408.13238}\ } (\bibinfo {year} {2024})}\BibitemShut {NoStop}%
\bibitem [{\citenamefont {Pouyez}\ \emph {et~al.}(2024{\natexlab{b}})\citenamefont {Pouyez}, \citenamefont {Grismayer}, \citenamefont {Grech},\ and\ \citenamefont {Riconda}}]{SuppMat}%
  \BibitemOpen
  \bibfield  {author} {\bibinfo {author} {\bibfnamefont {M.}~\bibnamefont {Pouyez}}, \bibinfo {author} {\bibfnamefont {T.}~\bibnamefont {Grismayer}}, \bibinfo {author} {\bibfnamefont {M.}~\bibnamefont {Grech}},\ and\ \bibinfo {author} {\bibfnamefont {C.}~\bibnamefont {Riconda}},\ }\bibfield  {title} {\bibinfo {title} {Supplemental material},\ }\href@noop {} {\  (\bibinfo {year} {2024}{\natexlab{b}})}\BibitemShut {NoStop}%
\bibitem [{\citenamefont {Del~Gaudio}\ \emph {et~al.}(2019)\citenamefont {Del~Gaudio}, \citenamefont {Grismayer}, \citenamefont {Fonseca}, \citenamefont {Mori},\ and\ \citenamefont {Silva}}]{del2019bright}%
  \BibitemOpen
  \bibfield  {author} {\bibinfo {author} {\bibfnamefont {F.}~\bibnamefont {Del~Gaudio}}, \bibinfo {author} {\bibfnamefont {T.}~\bibnamefont {Grismayer}}, \bibinfo {author} {\bibfnamefont {R.}~\bibnamefont {Fonseca}}, \bibinfo {author} {\bibfnamefont {W.}~\bibnamefont {Mori}},\ and\ \bibinfo {author} {\bibfnamefont {L.}~\bibnamefont {Silva}},\ }\bibfield  {title} {\bibinfo {title} {Bright $\gamma$ rays source and nonlinear breit-wheeler pairs in the collision of high density particle beams},\ }\href@noop {} {\bibfield  {journal} {\bibinfo  {journal} {Physical Review Accelerators and Beams}\ }\textbf {\bibinfo {volume} {22}},\ \bibinfo {pages} {023402} (\bibinfo {year} {2019})}\BibitemShut {NoStop}%
\bibitem [{\citenamefont {Mironov}\ \emph {et~al.}(2014)\citenamefont {Mironov}, \citenamefont {Narozhny},\ and\ \citenamefont {Fedotov}}]{mironov2014collapse}%
  \BibitemOpen
  \bibfield  {author} {\bibinfo {author} {\bibfnamefont {A.}~\bibnamefont {Mironov}}, \bibinfo {author} {\bibfnamefont {N.}~\bibnamefont {Narozhny}},\ and\ \bibinfo {author} {\bibfnamefont {A.}~\bibnamefont {Fedotov}},\ }\bibfield  {title} {\bibinfo {title} {Collapse and revival of electromagnetic cascades in focused intense laser pulses},\ }\href@noop {} {\bibfield  {journal} {\bibinfo  {journal} {Physics Letters A}\ }\textbf {\bibinfo {volume} {378}},\ \bibinfo {pages} {3254} (\bibinfo {year} {2014})}\BibitemShut {NoStop}%
\bibitem [{\citenamefont {Mercuri-Baron}\ \emph {et~al.}(2024)\citenamefont {Mercuri-Baron}, \citenamefont {Mironov}, \citenamefont {Riconda}, \citenamefont {Grassi},\ and\ \citenamefont {Grech}}]{mercuri2024growth}%
  \BibitemOpen
  \bibfield  {author} {\bibinfo {author} {\bibfnamefont {A.}~\bibnamefont {Mercuri-Baron}}, \bibinfo {author} {\bibfnamefont {A.}~\bibnamefont {Mironov}}, \bibinfo {author} {\bibfnamefont {C.}~\bibnamefont {Riconda}}, \bibinfo {author} {\bibfnamefont {A.}~\bibnamefont {Grassi}},\ and\ \bibinfo {author} {\bibfnamefont {M.}~\bibnamefont {Grech}},\ }\bibfield  {title} {\bibinfo {title} {Growth rate of self-sustained qed cascades induced by intense lasers},\ }\href@noop {} {\bibfield  {journal} {\bibinfo  {journal} {arXiv preprint arXiv:2402.04225}\ } (\bibinfo {year} {2024})}\BibitemShut {NoStop}%
\bibitem [{\citenamefont {Heitler}(1984)}]{heitler1984quantum}%
  \BibitemOpen
  \bibfield  {author} {\bibinfo {author} {\bibfnamefont {W.}~\bibnamefont {Heitler}},\ }\href@noop {} {\emph {\bibinfo {title} {The quantum theory of radiation}}}\ (\bibinfo  {publisher} {Courier Corporation},\ \bibinfo {year} {1984})\BibitemShut {NoStop}%
\bibitem [{\citenamefont {Selivanov}\ and\ \citenamefont {Fedotov}(2024)}]{selivanov2024final}%
  \BibitemOpen
  \bibfield  {author} {\bibinfo {author} {\bibfnamefont {Y.}~\bibnamefont {Selivanov}}\ and\ \bibinfo {author} {\bibfnamefont {A.}~\bibnamefont {Fedotov}},\ }\bibfield  {title} {\bibinfo {title} {Final multiplicity of a qed cascade in generalized heitler model},\ }\href@noop {} {\bibfield  {journal} {\bibinfo  {journal} {arXiv preprint arXiv:2408.06466}\ } (\bibinfo {year} {2024})}\BibitemShut {NoStop}%
\bibitem [{\citenamefont {Di~Piazza}\ \emph {et~al.}(2018)\citenamefont {Di~Piazza}, \citenamefont {Tamburini}, \citenamefont {Meuren},\ and\ \citenamefont {Keitel}}]{di2018implementing}%
  \BibitemOpen
  \bibfield  {author} {\bibinfo {author} {\bibfnamefont {A.}~\bibnamefont {Di~Piazza}}, \bibinfo {author} {\bibfnamefont {M.}~\bibnamefont {Tamburini}}, \bibinfo {author} {\bibfnamefont {S.}~\bibnamefont {Meuren}},\ and\ \bibinfo {author} {\bibfnamefont {C.}~\bibnamefont {Keitel}},\ }\bibfield  {title} {\bibinfo {title} {Implementing nonlinear compton scattering beyond the local-constant-field approximation},\ }\href@noop {} {\bibfield  {journal} {\bibinfo  {journal} {Physical Review A}\ }\textbf {\bibinfo {volume} {98}},\ \bibinfo {pages} {012134} (\bibinfo {year} {2018})}\BibitemShut {NoStop}%
\bibitem [{\citenamefont {Artemenko}\ \emph {et~al.}(2019)\citenamefont {Artemenko}, \citenamefont {Krygin}, \citenamefont {Serebryakov}, \citenamefont {Nerush},\ and\ \citenamefont {Kostyukov}}]{artemenkoPPCF2019}%
  \BibitemOpen
  \bibfield  {author} {\bibinfo {author} {\bibfnamefont {I.~I.}\ \bibnamefont {Artemenko}}, \bibinfo {author} {\bibfnamefont {M.~S.}\ \bibnamefont {Krygin}}, \bibinfo {author} {\bibfnamefont {D.~A.}\ \bibnamefont {Serebryakov}}, \bibinfo {author} {\bibfnamefont {E.~N.}\ \bibnamefont {Nerush}},\ and\ \bibinfo {author} {\bibfnamefont {I.~Y.}\ \bibnamefont {Kostyukov}},\ }\bibfield  {title} {\bibinfo {title} {Global constant field approximation for radiation reaction in collision of high-intensity laser pulse with electron beam},\ }\href {https://doi.org/10.1088/1361-6587/ab1712} {\bibfield  {journal} {\bibinfo  {journal} {Plasma Physics and Controlled Fusion}\ }\textbf {\bibinfo {volume} {61}},\ \bibinfo {pages} {074003} (\bibinfo {year} {2019})}\BibitemShut {NoStop}%
\bibitem [{\citenamefont {Zhang}\ \emph {et~al.}(2023)\citenamefont {Zhang}, \citenamefont {Grismayer},\ and\ \citenamefont {Silva}}]{zhang2023signatures}%
  \BibitemOpen
  \bibfield  {author} {\bibinfo {author} {\bibfnamefont {W.}~\bibnamefont {Zhang}}, \bibinfo {author} {\bibfnamefont {T.}~\bibnamefont {Grismayer}},\ and\ \bibinfo {author} {\bibfnamefont {L.}~\bibnamefont {Silva}},\ }\bibfield  {title} {\bibinfo {title} {Signatures for strong-field qed in the quantum limit of beamstrahlung},\ }\href@noop {} {\bibfield  {journal} {\bibinfo  {journal} {Physical Review A}\ }\textbf {\bibinfo {volume} {108}},\ \bibinfo {pages} {042816} (\bibinfo {year} {2023})}\BibitemShut {NoStop}%
\bibitem [{\citenamefont {Derouillat}\ \emph {et~al.}(2018)\citenamefont {Derouillat}, \citenamefont {Beck}, \citenamefont {Perez}, \citenamefont {Vinci}, \citenamefont {Chiaramello}, \citenamefont {Grassi}, \citenamefont {Flé}, \citenamefont {Bouchard}, \citenamefont {Plotnikov}, \citenamefont {Aunai}, \citenamefont {Dargent}, \citenamefont {Riconda},\ and\ \citenamefont {Grech}}]{smilei_derouillat_2018}%
  \BibitemOpen
  \bibfield  {author} {\bibinfo {author} {\bibfnamefont {J.}~\bibnamefont {Derouillat}}, \bibinfo {author} {\bibfnamefont {A.}~\bibnamefont {Beck}}, \bibinfo {author} {\bibfnamefont {F.}~\bibnamefont {Perez}}, \bibinfo {author} {\bibfnamefont {T.}~\bibnamefont {Vinci}}, \bibinfo {author} {\bibfnamefont {M.}~\bibnamefont {Chiaramello}}, \bibinfo {author} {\bibfnamefont {A.~D.}\ \bibnamefont {Grassi}}, \bibinfo {author} {\bibfnamefont {M.}~\bibnamefont {Flé}}, \bibinfo {author} {\bibfnamefont {G.}~\bibnamefont {Bouchard}}, \bibinfo {author} {\bibfnamefont {I.}~\bibnamefont {Plotnikov}}, \bibinfo {author} {\bibfnamefont {N.}~\bibnamefont {Aunai}}, \bibinfo {author} {\bibfnamefont {J.}~\bibnamefont {Dargent}}, \bibinfo {author} {\bibfnamefont {C.}~\bibnamefont {Riconda}},\ and\ \bibinfo {author} {\bibfnamefont {M.}~\bibnamefont {Grech}},\ }\bibfield  {title} {\bibinfo {title} {Smilei : A collaborative, open-source, multi-purpose particle-in-cell code for plasma simulation}\ }\href
  {https://doi.org/10.1016/J.CPC.2017.09.024} {10.1016/J.CPC.2017.09.024} (\bibinfo {year} {2018})\BibitemShut {NoStop}%
\end{thebibliography}
\end{document}

\title{Supplementary material for\\Kinetic structure of strong-field QED showers in crossed electromagnetic fields}

\author{M. Pouyez}
\affiliation{LULI, Sorbonne Université, CNRS, CEA, École Polytechnique, Institut Polytechnique de Paris, F-75255 Paris, France}
\author{T. Grismayer}
\affiliation{GoLP/Instituto de Plasmas e Fusão Nuclear, Instituto Superior Técnico, Universidade de Lisboa, 1049-001 Lisboa, Portugal}
\author{M. Grech}
\affiliation{LULI, CNRS, CEA, Sorbonne Universit\'{e}, École Polytechnique, Institut Polytechnique de Paris, F-91128 Palaiseau, France}
\author{C. Riconda}
\affiliation{LULI, Sorbonne Université, CNRS, CEA, École Polytechnique, Institut Polytechnique de Paris, F-75255 Paris, France}

\maketitle

\section{Generalites}

\subsection{Units and definition}

SI units and standard notations for physical constants will be used throughout the paper:
$c$ stands for the speed of light in vacuum, $m$ for the electron mass, $e$ for the elementary charge, $\alpha = e^2/(4\pi\varepsilon_0\hbar c)$ for the fine structure constant, $\varepsilon_0$ for the permittivity of vacuum and $\tau_c = \hbar/(m c^2)$ for the Compton time. The value $B_s=m_e^2c^2/(e \hbar)$ and $E_s=m_e^2c/(e \hbar)$ corresponds to the magnetic and electric Schwinger limit. The quantum parameters for leptons is define by $\chi=\gamma\sqrt{(\mathbf{E}+\mathbf{v}\times\mathbf{B})^2-(\mathbf{v}\cdot\mathbf{E})^2/c^2}/E_s$ with $\mathbf{B}$ and $\mathbf{E}$ the magnetic and electric field vector, $\gamma$ the Lorentz factor and $v$ the velocity of electrons. In a constant and uniform magnetic field perpendicular to the initial direction of the lepton, the quantum parameter simply writes $\chi=\sqrt{\gamma^2-1}B/B_s$. Similarly, for photons, we have $\chi_\gamma=\gamma_\gamma B/B_s$.

\subsection{Photon emission and pair production rates}\label{sec:rateDef}

The processes of high-energy photon emission (nonlinear inverse Compton scattering)
and electron-positron pair production (Breit-Wheeler process)
are described by their (energy) differential rates computed in the locally constant cross-field approximation
(LCFA):
\begin{eqnarray}
    \label{eq:w-nics}w(\gamma,\gamma_\gamma) &=& \left.\frac{d\Wcs}{d\gamma_\gamma}\right\vert_\chi = \frac{\alpha}{\tau_c}\,\frac{Q(\chi,\gamma_\gamma/\gamma)}{\gamma^2}\,,\\
    \label{eq:w-bw}\overline{w}(\gamma_\gamma,\gamma) &=& \left.\frac{d\Wbw}{d\gamma}\right\vert_{\chi_\gamma} = \frac{\alpha}{\tau_c}\,\frac{\overline{Q}(\chi_\gamma,\gamma/\gamma_\gamma)}{\gamma_\gamma^2}\,,
\end{eqnarray}
where $w_\chi(\gamma,\gamma_\gamma)\,dt\,d\gamma_\gamma$ denotes the probability for an electron (equivalently positron) of energy $\gamma\,m c^2$ 
and quantum parameter $\chi$ to emit a high-energy photon with energy in between $\gamma_\gamma mc^2$ and $(\gamma_\gamma+d\gamma_\gamma) mc^2$ between times $t$ and $t+dt$,
while $\overline{w}_{\chi_\gamma}(\gamma_\gamma,\gamma)\,dt\,d\gamma$ denotes that of a photon to decay into a pair with the electron carrying an energy 
in between $\gamma mc^2$ and $(\gamma+d\gamma) mc^2$. 
The functions $F(\chi,\xi)$ and $G(\chi_\gamma,\zeta)$ are 
\begin{eqnarray}
    \label{eq:F-nics} Q(\chi,\xi) &=& \frac{1}{\pi\sqrt{3}}\,\left[\frac{\xi^2}{1-\xi}\,K_{2/3}(\nu) + \int_\nu^{\infty}K_{5/3}(y)dy\right]\,,\\
    \label{eq:G-bw} \overline{Q}(\chi_\gamma,\zeta) &=& \frac{1}{\pi\sqrt{3}}\,\left[\frac{1}{\zeta(1-\zeta)}\,K_{2/3}(\mu) - \int_\mu^{\infty}K_{5/3}(y)dy\right]\,,
\end{eqnarray}
with $\xi=\gamma_\gamma/\gamma$, $\nu=2\xi/[3\chi(1-\xi)]$, $\zeta=\gamma/\gamma_\gamma$ and $\mu=2/[3\chi_\gamma\zeta(1-\zeta)]$,
and where $K_n(x)$ denotes the modified Bessel function of the second kind.\\

Integrating these differential rates over all possible daughter particle energy 
[photons for Eq.~\eqref{eq:w-nics} and electrons/positrons for Eq.~\eqref{eq:w-bw}] leads to the photon emission and pair production rates
\begin{eqnarray}
    \label{eq:W-nics} \Wcs(\gamma,\chi) = \frac{\alpha}{\tau_c}\,\frac{a(\chi)}{\gamma}\,,\\
    \label{eq:W-bw} \Wbw(\gamma_\gamma,\chi_\gamma) = \frac{\alpha}{\tau_c}\,\frac{b(\chi_\gamma)}{\gamma_\gamma}\,,
\end{eqnarray}
where the functions $a(\chi)$ and $b(\chi_\gamma)$ are 
\begin{eqnarray}
    \label{eq:a-nics} a(\chi)\!&=&\!\int_0^1 d\xi\,F(\chi,\xi) \longrightarrow 
    \begin{cases}
        a_1\,\chi, & {\rm for}~\chi \ll 1\\
        a_2\,\chi^{2/3}, & {\rm for}~\chi \gg 1
    \end{cases} \,,\\
    \label{eq:b-bw} b(\chi_\gamma)\!&=&\!\int_0^1 d\xi\,G(\chi_\gamma,\xi) \longrightarrow
    \begin{cases}
        b_1\,\chi_\gamma\,e^{-\frac{8}{3\chi_\gamma}}, & {\rm for}~\chi_\gamma \ll 1\\
        b_2\,\chi_\gamma^{2/3}, & {\rm for}~\chi_\gamma \gg 1
    \end{cases} \,.
\end{eqnarray}
with $a_1=5\sqrt{3}/6\simeq1.44$, $a_2=14\Gamma(2/3)/3^{7/3}\simeq 1.46$, $b_1=3/16(3/2)^{1/2}\simeq0.23$, $b_2=3^{5/3}5\Gamma^4(2/3)/(28\pi^2)\simeq0.38$.

\subsection{Radiation time: derivation of Eq. (3) of the manuscript}
The radiation time corresponds to the necessary time for an electron to lose a significant amount of energy, ie to emit a photon at the average energy \cite{akhiezer1994kinetic}:

\begin{eqnarray}
    T_r= \int_1^{\gamma_0}\frac{d\gamma}{\int_0^\infty  d\gamma_\gamma \gamma_\gamma w(\gamma,\gamma_\gamma)}.
\end{eqnarray}
In the limit of $\chi\gg 1$ and $\gamma\gg1$ we have:
\begin{eqnarray}
    \frac{\int_0^\infty  d\gamma_\gamma \gamma_\gamma w(\gamma,\gamma_\gamma)}{\int_0^\infty  d\gamma_\gamma w(\gamma,\gamma_\gamma)}= \frac{16}{63}\gamma
\end{eqnarray}
corresponding to the average energy of a photon emitted by an electron at $\gamma$. We also have:
\begin{eqnarray}
    \int_0^\infty  d\gamma_\gamma w(\gamma,\gamma_\gamma)\simeq a_2\frac{\alpha}{\tau_c}\,\frac{\chi^{2/3}}{\gamma}
    =a_2\frac{\alpha}{\tau_c}\,\left(\frac{B}{B_s}\right)^{2/3}\gamma^{-1/3}.
\end{eqnarray}
Injecting into the definition of $T_r$ we obtain:
\begin{eqnarray}
    T_{r} \equiv c_1\,\frac{\tau_c}{\alpha}\,\gamma_0^{1/3}\left(\frac{B_s}{B}\right)^{2/3} \equiv c_1 \,\frac{\tau_c}{\alpha}\gamma_0\chi_0^{-2/3} \,
\end{eqnarray}
where $c_1=3^{16/3}/[32\Gamma(2/3)]\simeq 8.09$.

\subsection{Kinetic equations: Eqs. (1)-(2) of the manuscript}

As presented in the manuscript we split each species into successive generations: a lepton of generation $(n)$ emits a photon of generation $(n)$ that can in turn decay into leptons of generation $(n+1)$. We assume that all particles participating in high-energy photon emission and pair production are ultra-relativistic ($\gamma \gg 1$) at all times. This allows us to approximate electron and positron velocity as that of light and
to consider that high-energy photons and electron-positron pairs are emitted/created with a momentum aligned with that of the particle they originate from. We consider that the field created by the charged particles is negligible in front of the external magnetic field and finally, we describe the two processes by their energy differential rates computed in the locally constant cross-field approximation (LCFA) (Sec. \ref{sec:rateDef}).

With these approximations, we obtain the following set of equations:
\begin{eqnarray}
    \nonumber \partial_t f_\pm^{(n)}(\gamma,t) &=& \int_0^{\infty}\!\! d\gamma_\gamma\, w(\gamma+\gamma_\gamma,\gamma_\gamma)\,f_\pm^{(n)}(\gamma+\gamma_\gamma,t)
    - \int_0^{\infty}\!\! d\gamma_\gamma\, w(\gamma,\gamma_\gamma)\,f_\pm^{(n)}(\gamma,t) \nonumber\\
    &+&\int_0^{\infty}\!\! d\gamma_\gamma\, \overline{w}(\gamma_\gamma,\gamma)\,f_\gamma^{(n-1)}(\gamma_\gamma,t)\,, \label{eq:fpn}\\
     \partial_t f_\gamma^{(n)}(\gamma_\gamma,t) &=& \int_1^{\infty}\!\! d\gamma\, w(\gamma,\gamma_\gamma)\,\left[f_-^{(n)}(\gamma,t)+f_+^{(n)}(\gamma,t)\right]
    - \Wbw(\gamma_\gamma)f_\gamma^{(n)}(\gamma_\gamma,t) \, . \label{eq:fgn}
\end{eqnarray}

Equation~\eqref{eq:fgn} admits a general solution of the form:
\begin{eqnarray}\label{eq:gensol-fgn}
    f_\gamma^{(n)}(\gamma,t) = \int_1^\infty \!\!d\gamma\,w(\gamma,\gamma_\gamma)\,\int_0^t dt'\,\left[f_-^{(n)}(\gamma,t')+f_+^{(n)}(\gamma,t')\right]\,e^{-\overline{W}(\gamma_\gamma)\,(t-t')}.
\end{eqnarray}
In contrast, it is not possible to give a general solution for Eq.~\eqref{eq:fpn}. Nevertheless, integrating it over all energies, the first two terms give a vanishing contribution (photon emission conserves the number of leptons). Remembering that the magnetic field is constant, we obtain - using Eq.~\eqref{eq:gensol-fgn} and after some algebra - the number of pairs:
\begin{eqnarray}\label{eq:Nn}
    N_\pm^{(n)}(t) &=& \int_0^\infty\!d\gamma_\gamma
    \int_0^t\!dt' \int_1^\infty \!d\gamma \left[f_{-}^{(n-1)}(\gamma,t')+f_{+}^{(n-1)} (\gamma,t')\right]
     w(\gamma,\gamma_\gamma)\,
    \left[1-e^{-(t-t')\, \Wbw(\gamma_\gamma)}\right]\ \, \nonumber \\ 
    &+& \int_0^\infty\!d\gamma_\gamma f_\gamma^{(n-1)}(\gamma_\gamma,t=0)\left[1-e^{-t\, \Wbw(\gamma_\gamma)}\right]\,.
\end{eqnarray}
The first term of this equation corresponds to the convolution of the spectrum of photons emitted by the lepton of generation $(n-1)$ by the photon decay probability. The second term corresponds to the pairs generated by the initial photons, and this second term is non-zero  only for the photon seeding case for which $f_\gamma^{(0)}(\gamma_\gamma,t=0) \ne 0$.

\section{Calculation for short times}

\subsection{General solution: derivation of Eqs.~(4)-(9) of the manuscript}

Over short time scales ($t\ll T_r$), the leptons of generation $(n)$ have not interacted enough with the field to lose a significant part of their energy and we can neglect the radiation operator in the dynamics of leptons. The system to solve simply write:
\begin{eqnarray}
     \partial_t f_\pm^{(n)}(\gamma,t) &=&\int_0^{\infty}\!\! d\gamma_\gamma\, \overline{w}(\gamma_\gamma,\gamma)\,f_\gamma^{(n-1)}(\gamma_\gamma,t)\,, \label{eq:fpn:sh}\\
    \partial_t f_\gamma^{(n)}(\gamma_\gamma,t) &=& \int_1^{\infty}\!\! d\gamma\, w(\gamma,\gamma_\gamma)\,\left[f_-^{(n)}(\gamma,t)+f_+^{(n)}(\gamma,t) \right]
    - \Wbw(\gamma_\gamma)f_\gamma^{(n)}(\gamma_\gamma,t) \, . \label{eq:fgn:sh}
\end{eqnarray}
Where the solutions are given by:
\begin{eqnarray}
    f_\pm^{(n)}(\gamma,t)&=&\int_0^\infty d\gamma_\gamma \overline{w}(\gamma_\gamma,\gamma)\int_0^t dt' f_\gamma^{(n-1)}(\gamma_\gamma,t'), \\
    f_\gamma^{(n)}(\gamma_\gamma,t)&=&e^{-\Wbw(\gamma_\gamma)t}\int_1^\infty d\gamma w(\gamma,\gamma_\gamma)\int_0^t dt' e^{\Wbw(\gamma_\gamma)t'}\left[f_+^{(n)}(\gamma,t')+f_-^{(n)}(\gamma,t')\right].
\end{eqnarray}
Considering $t\ll T_r$ meaning that $\forall \gamma$, $\Wcs(\gamma)t \ll 1$ we also have $\forall \gamma_\gamma$, $\Wbw(\gamma_\gamma)t\ll 1$ because $\forall x$, $\Wcs(x)>\Wbw(x)$. We are now in position to consider that the probability for a photon to decay is low and then $\exp{(-\Wbw(\gamma_\gamma)t)}\simeq 1-\Wbw(\gamma_\gamma)t+o(t)$. It follows that $f_-^{(0)}(\gamma,t)= f_-^{(0)}(\gamma,t=0)$ and the solutions writes:
\begin{eqnarray}
    f_\gamma^{(n-1)}(\gamma_\gamma,t)&=& \frac{2^{n-1}}{(2n-1)!}G^{(n-1)}(\gamma_\gamma) \left(\frac{t}{T_r}\right)^{2n-1} + o\left(\left[\frac{t}{T_r}\right]^{2n-1}\right) ,\label{eq:sh:sol:fgn} \\
    f_\pm^{(n)}(\gamma,t)&=& \frac{2^{n-1}}{(2n)!}L^{(n)}(\gamma)\left(\frac{t}{T_r}\right)^{2n}+ o\left(\left[\frac{t}{T_r}\right]^{2n}\right) .\label{eq:sh:sol:fpn}
\end{eqnarray}
with
\begin{eqnarray}
    G^{(n)}(\gamma_\gamma)&=& \int_1^{\infty}\!\! d\gamma \, T_r w(\gamma,\gamma_\gamma)L^{(n)}(\gamma) ,\label{eq:sh:Gn} \\
    L^{(n)}(\gamma)&=& \int_0^{\infty}\!\! d\gamma_\gamma  \, T_r \overline{w}(\gamma_\gamma,\gamma)G^{(n-1)}(\gamma_\gamma), \label{eq:sh:Ln}\\
    L^{(0)}(\gamma)&=&f_-^{(0)}(\gamma,t=0).
\end{eqnarray}
The first generation dominated the pair production at short times and the number of total pairs can be estimated by the first pairs generation only. Thus, for $N_0$ incident electron at $\gamma_0$ the shower multiplicity $N_\pm/N_0$ write:
\begin{eqnarray}
    N_\pm(t)/N_0=\left(\frac{t}{T_r}\right)^2\frac{T_r^2}{2}\int_0^{\gamma_0} d\gamma_\gamma \overline{W}(\gamma_\gamma)w(\gamma_0,\gamma_\gamma)\longrightarrow \left(\frac{t}{T_r}\right)^2 \times
    \begin{cases}
        1.25\chi_0^{2/3}e^{-\frac{16}{3\chi_0}}, & {\rm for}~\chi_0 \lesssim 1\\
        6.1\ln(\chi_0)-13.6, & {\rm for}~\chi_0 \gg 1
    \end{cases} \,.\label{eq:st:N1exact}
\end{eqnarray}

\subsection{Estimate with reduced rates}
In the literature \cite{akhiezer1994kinetic,selivanov2024final} we can find some solutions for the average rate. In this section, we show how to link the general solution with the average rate approximation. The rates in the limits of $\chi,\chi_\gamma \gg 1$ can be estimated as:
\begin{eqnarray}
    w(\gamma,\gamma_\gamma)&\simeq& \Wcs(\gamma)\delta(\gamma/4-\gamma_\gamma)\simeq a_2\frac{\alpha}{\tau_c}\frac{\chi^{2/3}}{\gamma}\delta(\gamma/4-\gamma_\gamma),\\
    \overline{w}(\gamma_\gamma,\gamma)&\simeq& \frac{1}{2}\Wbw(\gamma_\gamma)\left[\delta(\gamma-\gamma_\gamma)+\delta(\gamma)\right]\simeq \frac{1}{2}b_2\frac{\alpha}{\tau_c}\frac{\chi_\gamma^{2/3}}{\gamma_\gamma}\left[\delta(\gamma-\gamma_\gamma )+\delta(\gamma)\right], \label{approx_rate_BW}
\end{eqnarray}
where we have considered that when a lepton of energy $\gamma$ emits a photon, this one is at the average energy ie at $16/63 \, \gamma\simeq \gamma/4$ and when a photon decays into pairs it gives all his energy to one lepton and nothing to the other one. To simplify the calculation we consider that the particle with all the energy is always the positron. This assumption does not influence the number of pairs. It follow that the solution \eqref{eq:sh:sol:fgn} and \eqref{eq:sh:sol:fpn} can be rewrite as:
\begin{eqnarray}
    f_\gamma^{(n-1)}(\gamma_\gamma,t)&=& \frac{1}{(2n-1)!}\widetilde{G}^{(n-1)}(\gamma_\gamma) \left(\frac{t}{T_r}\right)^{2n-1} + o\left(\left[\frac{t}{T_r}\right]^{2n-1}\right), \\
    f_+^{(n)}(\gamma,t)&=& \frac{1}{(2n)!}\widetilde{L}^{(n)}(\gamma)\left(\frac{t}{T_r}\right)^{2n}+ o\left(\left[\frac{t}{T_r}\right]^{2n}\right)  .
\end{eqnarray}
where 
\begin{eqnarray}
    \widetilde{G}^{(n)}(\gamma_\gamma)&=& 4^{n+1}T_r^{2n+1}\delta(4^{n+1}\gamma_\gamma-\gamma_0)\prod_{i=0}^n \Wcs(4^{i+1}\gamma_\gamma) \prod_{i=1}^n\Wbw(4^i\gamma_\gamma), \\
    \widetilde{L}^{(n)}(\gamma)&=&  4^{n}T_r^{2n}\delta(4^{n}\gamma-\gamma_0)\prod_{i=1}^n \Wcs(4^{i}\gamma) \prod_{i=1}^n\Wbw(4^{i-1}\gamma).
\end{eqnarray}
Integrating $f_+^{(n)}$ over all energies we obtain the number of pairs:
\begin{eqnarray}
    N_\pm^{(n)}(t)=\frac{4^n T_r^{2n}}{(2n)!}\left(\frac{t}{T_r}\right)^{2n} \prod_{i=1}^{n}\Wbw\left(\frac{\gamma_0}{4^i}\right)\Wcs\left(\frac{\gamma_0}{4^{i-1}}\right).
\end{eqnarray}
Finally using the expression of the rate in the limits $\chi,\chi_\gamma\gg1$ we obtain:
\begin{eqnarray}\label{eq:sh:sol:Nn}
    N_\pm^{(n)}(t)=2^{2n^2/3} \left(\frac{189}{16}\sqrt{\frac{b_2}{a_2}}\right)^{2n}\frac{1}{(2n)!} \left(\frac{t}{T_r}\right)^{2n} + o\left(\left[\frac{t}{T_r}\right]^{2n}\right)
\end{eqnarray}
where $189/16\sqrt{b_2/a_2}\simeq6.026$. These solutions give a good estimate for the number of pairs but the full integration of Eq. \eqref{eq:st:N1exact} should be used for a precise prediction.

\section{Calculation for the long time scale}

\subsection{Photon decay probability: derivation of Eq. (13) of the manuscript}\label{sec:approxProba}

\begin{figure}
    \centering
    \includegraphics[width=0.4\linewidth]{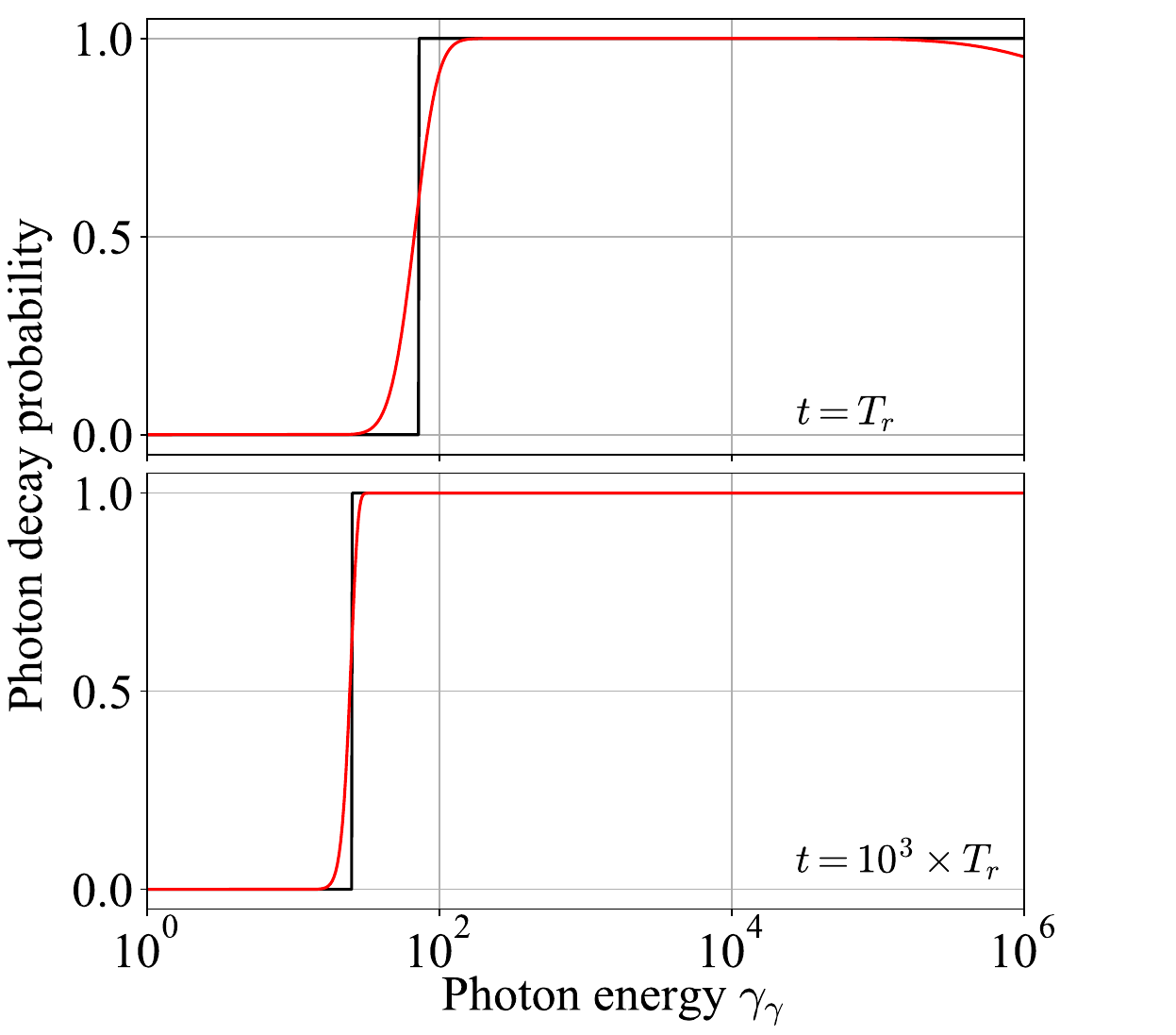}
    \caption{Photon decay probability as a function of photon energy. On the top panel the probability after $t=T_r$ and on the bottom one after $t=10^3T_r$ with $T_r$ calculated using $ B/B_s=0.01$ and $\gamma_0=10^6$. In black solid line the approximation $P(\gamma_\gamma,t)=\Theta(\gamma_\gamma-\gamma_l(t)$ and in red solid line the exact calculation $P(\gamma_\gamma,t)=1-e^{-t\Wbw(\gamma_\gamma,t)}$. }
    \label{fig:prob}
\end{figure}

In a long time scale ($t\gg T_r$) we can consider that the majority of the surviving photons admit low quantum parameters because all photons emitted at high energy have already decayed into pairs. Thus we consider that the probability of a photon which has interacted with the field during a time $t$ is equivalent to an Heaviside function evaluated at $\gamma_l(t)$: $P(\gamma,t)=1-e^{-t\Wbw(\gamma_\gamma,t)}\simeq \Theta (\gamma_\gamma-\gamma_l(t))$. This $\gamma_l$ physically corresponds to the minimal energy for a photon needed to decay and should, in our condition, correspond to a low quantum parameter. We define this quantity by the solution of $t \Wbw(\gamma_l(t))=1$, a condition meaning that the probability of a photon to decay is equal to $0.63$. Using the asymptotic expression of $\Wbw$ function we have:
\begin{eqnarray}
    \gamma_l(t)=\frac{8B_s}{3B }\ln^{-1}\left(b_1\alpha\frac{t}{\tau_c}\frac{B}{B_s}\right)=\frac{8B_s}{3B }\ln^{-1}\left(c_2\frac{t}{T_r}\chi_0^{1/3}\right)\, . 
\end{eqnarray}
where $c_2=c_1 \, 3/16(3/2)^{1/2}\simeq1.861$. In the next we denote $\chi_l(t)=B/B_s  \gamma_l(t)$. A comparison of the Heaviside approximation (black solid line $\Theta(\gamma_\gamma-\gamma_l(t)$) and the exact photon decay probability (red solid line $1-e^{-t\Wbw(\gamma_\gamma,t)}$) is given in Fig. \ref{fig:prob} as a function of photon energy for two different times of interaction. As shown, the Heaviside function well represents the photon decay probability for $t\gg T_r$. 

\subsection{Recurrence relation for the photon spectrum: derivation of Eqs. (10)-(12),(14) of the manuscript. }

In the long time limit ($t \gg T_r$), the leptons have already lost most of their energy through radiation, and mainly low energy photons remain that slowly decay through their interaction with the magnetic field. When computing the number of pairs $N_\pm^{(n)}(t)$,
one can thus consider that the photons from which the pairs originate have been created at a time $t' \ll t$ and formally take $t'=0$ in Eq.~\eqref{eq:Nn}. This leads:
\begin{eqnarray}\label{eq:Npm-longtime}
   N_{\pm}^{(n)}(t) = \int_0^{\infty}\!\!d\gamma_\gamma\,F_{\gamma}^{(n-1)}(\gamma_\gamma,t)\,P(\gamma_\gamma,t)\,,
\end{eqnarray}
where we have introduced:
\begin{eqnarray}\label{eq:Fnm1}
    F^{(n-1)}_\gamma(\gamma_\gamma,t) = \int_0^t\!dt'\!\int_1^\infty\!\!d\gamma\,w(\gamma,\gamma_\gamma)\,\left[f_-^{(n-1)}(\gamma,t')+f_+^{(n-1)}(\gamma,t')\right]\,.
\end{eqnarray}
Recognizing that $F^{(n-1)}_\gamma(\gamma_\gamma,t)$ satisfies Eq.~\eqref{eq:fgn} without the second term (photon decay) in its right-hand-side, $F_\gamma(\gamma_\gamma,t)$ denotes the energy distribution of all emitted photons of the previous ($n-1$) generation. 

In order to compute the spectrum of pairs $f^{(n-1)}_\pm(\gamma,t')$ and obtain a recursive form for Eq. \eqref{eq:Fnm1}, we assume that the average energy is sufficient \cite{pouyez2024multiplicity} to estimate the photon spectrum emitted by the leptons. The energy at time $t>\tcr$ of one particle created at a time $\tcr$ with energy $\gcr$ is noted $\gamma(t,\gcr,\tcr)$ and is compute with it's equation of evolution \cite{niel2018pre}:
\begin{eqnarray}\label{eq:AvEnergy}
    \frac{d}{dt} \gamma(t,\gcr,\tcr) \simeq -\frac{2}{3}\frac{\alpha}{\tau_c}\,\chi^2 \, g\!\left(\chi\right)
\end{eqnarray}
where $g(\chi)$ is the so-called Gaunt factor of quantum radiation reaction. By introducing the function $\psi^{(n-1)}_\pm(\tcr,\gcr)$ to describe the pairs created at time $\tcr$ with energy $\gcr$ by photons of generation $(n-1)$, i.e. the creation term of Eq. \eqref{eq:fpn}, we have:

\begin{eqnarray}
    f_\pm^{(n-1)}(\gamma,t) = \int_0^{t}d\tcr \int_1^{\infty}d\gcr \psi^{(n-1)}_\pm(\tcr,\gcr) \delta(\gamma -  \gamma(t,\gcr,\tcr) ).
\end{eqnarray}

\noindent Since most of the leptons still contributing to the shower are created at $\tcr \ll t$ in this long-time regime, we neglect the explicit dependence on $\tcr$ in $ \gamma(t,\gcr,\tcr)$, and  obtain:

\begin{eqnarray}\label{eq:def_f_n:approx}
    f_\pm^{(n-1)}(\gamma,t) \simeq \int_1^{\infty}d\gcr \delta(\gamma -  \gamma(t,\gcr) ) \varphi^{(n-1)}(\gcr,t),
\end{eqnarray}
where we have introduce $\varphi^{(n-1)}(\gcr,t)=\int_0^{t}d\tcr \psi^{(n-1)}_\pm(\tcr,\gcr)$ the total energy distribution of $n-1$ leptons created at energy $\gcr$ before time $t$. Injecting Eq. \eqref{eq:def_f_n:approx} in Eq.~\eqref{eq:Fnm1}, and assuming that, in this long-time limit,  $\varphi^{(n-1)}_\pm\!\left(\gcr,t\right)$ varies very slowly in time (all pairs have been created long ago), we get an approximate form for $F_\gamma^{(n-1)}(\gamma_\gamma,t)$:
\begin{eqnarray}\label{eq:Fnm1-inter1}
    F^{(n-1)}_\gamma(\gamma_\gamma,t) = \int_1^{\infty}d\gcr\,\,\left[\varphi^{(n-1)}_-(\gcr,t)+\varphi^{(n-1)}_+(\gcr,t)\right]\,\,\int_{0}^t\!\!dt'\,w\!\left( \gamma(t',\gcr) ,\gamma_\gamma\right).
\end{eqnarray}
Further assuming that we look at very late times over which the leptons have radiated all their initial energy into photons and replacing the integral over time as an integral over energies, we finally get:
\begin{eqnarray}\label{eq:Fgnm1-new}
    F^{(n-1)}_\gamma(\gamma_\gamma,t) = \int_1^\infty d\gcr\,\,\left[\varphi^{(n-1)}_-(\gcr,t)+\varphi^{(n-1)}_+(\gcr,t)\right]\,\,I_\gamma\!\left(\gcr,\gamma_\gamma\right)\,,
\end{eqnarray}

\noindent where we have introduced the transfer function:
\begin{eqnarray}\label{eq:lt:Igamma}
    I_\gamma\left(\gcr,\gamma_\gamma\right) = \frac{3}{2}\frac{\tau_c}{\alpha} \int_1^{\gcr}\!\!d\gamma\,\frac{w(\gamma,\gamma_\gamma)}{\chi^2\,g(\chi)}\,,\quad{\rm with}\quad \chi=\gamma\,B/B_s\,.
\end{eqnarray}
This transfer function returns the final energy distribution of all photons emitted by a lepton created at an energy $\gcr$.
Thus, equation~\eqref{eq:Fgnm1-new} can be understood as follows, the generation $(n-2)$ of photons create pairs according to the distribution function $\varphi^{(n-1)}_\pm(\gcr,t)$ and each of these pairs creates photons according to the transfer function $I_\gamma(\gamma_\gamma,\gamma)$. By construction the function $\varphi^{(n-1)}_\pm(\gcr,t)$ satisfies Eq.~\eqref{eq:fpn} (for generation $n-1$) without accounting for the radiation reaction terms:

\begin{eqnarray}
    \partial_t \varphi^{(n-1)}_\pm(\gcr,t) = \int_0^\infty\!\!d\gamma_\gamma\,\overline{w}(\gamma_\gamma,\gcr)\,f_\gamma^{(n-2)}(\gamma_\gamma,t)\,.
\end{eqnarray}
Integrating in time and using Eq.~\eqref{eq:gensol-fgn}, we get after some algebra:
\begin{eqnarray}
    \varphi^{(n-1)}_\pm(\gcr,t) = \int_0^\infty\!\!\!d\gamma_\gamma\,\frac{\overline{w}(\gamma_\gamma,\gcr)}{\overline{W}(\gamma_\gamma)}
    \int_1^\infty \!\!\!d\gamma\,w\!\left(\gamma,\gamma_\gamma\right)
    \int_0^t\!\!dt' \left[f^{(n-2)}_-(\gamma,t')+f^{(n-2)}_+(\gamma,t')\right]\,
    \Big[1-\exp\left(-\overline{W}(\gamma_\gamma)(t-t')\right)\Big]\,.
\end{eqnarray}
We see that $\varphi^{(n-1)}_\pm(\gcr,t)$ depends on the distribution of leptons of the previous ($n-2$) generation. As we have done earlier to derive Eq.~\eqref{eq:Npm-longtime}, we now consider that these leptons were created at a time $t'\ll t$, so that we can formally take $t'=0$ in the previous equation. Using Eq.~\eqref{eq:Fnm1}, we then obtain:
\begin{eqnarray}
    \varphi^{(n-1)}_\pm(\gcr,t) = \int_0^\infty\!\!\!d\gamma_\gamma\,\frac{\overline{w}(\gamma_\gamma,\gcr)}{\overline{W}(\gamma_\gamma)}\,
    F_\gamma^{(n-2)}(\gamma_\gamma,t)\,
    P(\gamma_\gamma,t)\,.
\end{eqnarray}
Injecting in Eq.\eqref{eq:Fnm1-inter1} leads the recursive relation:
\begin{eqnarray}
    F_\gamma^{(n-1)}(\gamma_\gamma,t) = 2\int d\gcr\,I_\gamma\!\left(\gcr,\gamma_\gamma\right)\, \int_0^\infty\!\!\!d\gamma_\gamma'\,\frac{\overline{w}(\gamma_\gamma',\gcr)}{\overline{W}(\gamma_\gamma')}\,
    F_\gamma^{(n-2)}(\gamma_\gamma',t)\,
    P(\gamma_\gamma',t)\,,
\end{eqnarray}
that is Eq.~(12) of the Manuscript. Following Sec.~\ref{sec:approxProba}, i.e. using $P(\gamma_\gamma',t) \simeq \Theta(\gamma_\gamma'-\gamma_l(t))$,
and considering that, in the long-time limit,
photons decay into two pairs with identical energies [i.e. taking $\overline{w}(\gamma_\gamma',\gcr) \simeq \overline{W}(\gamma_\gamma')\,\delta(\gcr - \gamma_\gamma'/2)$], we finally get:
\begin{eqnarray}
    F_\gamma^{(n-1)}(\gamma_\gamma,t) = 2 \int_{\gamma_l(t)}^\infty\!\!\!d\gamma_\gamma'\,\,
    F_\gamma^{(n-2)}(\gamma_\gamma',t)\,\,
    I_\gamma\!\left(\gamma_\gamma'/2,\gamma_\gamma\right)\,\,.
\end{eqnarray}
This form, Eq.~(14) of the Manuscript, is in excellent agreement with Monte Carlo simulations as shown in Fig.~1(d) of the Manuscript.

\subsection{Calculation for the number of pairs: derivation of Eqs. (15)-(17) of the manuscript}
We are now in a position to obtain a recursive form for $N_\pm^{(n+1)}$ and $N_\pm^{(n)}$. We have:
\begin{eqnarray}
    N_\pm^{(n+1)}(t)&=&2 \int_{\gamma_l(t)}^{\infty}d\gamma_\gamma \int_{\gamma_l(t)}^{\infty} d\gamma I_\gamma(\gamma_\gamma,\gamma/2)F_\gamma^{(n-1)}(\gamma,t) \nonumber \\ 
    &=& 2\int_{2\gamma_l(t)}^{\infty} d\gamma F_\gamma^{(n-1)}(\gamma,t) J_\gamma(\gamma,\gamma_l) \label{eq:rec:approx1:Nn}
\end{eqnarray}
with $J_\gamma(\gamma,\gamma_l)=\int_{\gamma_l}^{\gamma/2}d\gamma_\gamma I_\gamma(\gamma_\gamma,\gamma/2)$. The lower integration bound in Eq. \eqref{eq:rec:approx1:Nn} is now $2\gamma_l(t)$ since $J_\gamma(\gamma,\gamma_l)$ is non zero only if $\gamma/2>\gamma_l(t)$. Integrating by part and defining $\widetilde{N}^{(n)}_\pm(\gamma,t)=\int_\gamma^{\infty} d\gamma_\gamma F^{(n-1)}(\gamma_\gamma,t)$ and $J_\gamma'(\gamma,\gamma_l)=d J(\gamma,\gamma_l)/d\gamma$ we obtain:
\begin{eqnarray}\label{eq:rec:before_laplace}
    N_\pm^{(n+1)}(t)=2\int_{2\gamma_l(t)}^{\infty}d\gamma \widetilde{N}_\pm^{(n)}(\gamma,t)J_\gamma'(\gamma,\gamma_l(t)).
\end{eqnarray}
The function $J_\gamma'$ admits a sharp maximum at $\gamma_m(\chi_l)$. We can thus use with very good approximation the Laplace method for the integration, leading to:

\begin{eqnarray}\label{eq:rec:approx2:Nn}
    N_\pm^{(n+1)}(t)\simeq 2\sqrt{2\pi}\widetilde{N}_\pm^{(n)}(\gamma_m(\gamma_l),t)J_\gamma'(\gamma_m(\gamma_l),\gamma_l)^{3/2}\lvert J_\gamma'''(\gamma_m(\gamma_l),\gamma_l) \rvert^{-1/2}.
\end{eqnarray}
The goal now is to obtain an estimate for the function $J_\gamma'$ by considering the asymptotic expression of the low energetic part of the non-linear Compton Scattering differential rate, given by:
\begin{eqnarray}
    w(\gamma_\gamma,\gamma)\simeq k_1\frac{\alpha}{\tau_c}\left(\frac{B }{B_s}\right)^{2/3}\gamma_\gamma^{-2/3}\gamma^{-2/3} \Theta(\gamma-\gamma_\gamma)
\end{eqnarray}
where $k_1=\Gamma(-1/3)/[3^{1/3}\Gamma(-2/3)\Gamma(2/3)]\simeq 0.517$. By taking the asymptotic value of the Gaunt factor at $\chi\gg1$ \cite{baier1998electromagnetic},  $g(\chi)\simeq k_2\chi^{-4/3}$,  with $k_2=2.44^{-2/3}$ we obtain: 
\begin{eqnarray}
    I_\gamma(\gamma_\gamma,\gamma)&\simeq& \frac{9k_1}{2k_2}\gamma_\gamma^{-1}\left[1-\left(\frac{\gamma_\gamma}{\gamma}\right)^{1/3}\right]\Theta(\gamma-\gamma_\gamma), \\
    J_\gamma'(\chi,\chi_l)&=&\frac{9k_1}{2k_2}\gamma^{-1}\left[1-2^{1/3}\left(\frac{\gamma_l}{\gamma}\right)^{1/3}\right]\Theta (\gamma-2\gamma_l).
\end{eqnarray}
The maximum of $J_\gamma'$ is located at $\gamma_m=128\gamma_l/27$ so equation Eq. \eqref{eq:rec:approx2:Nn} become:
\begin{eqnarray}
    N_\pm^{(n+1)}(t)= \frac{9}{8}\sqrt{6\pi}\frac{k_1}{k_2}\widetilde{N}^{(n)}_\pm\left(\gamma_l(t)\frac{128}{27},t\right)
\end{eqnarray}
which gives the simple solution:
\begin{eqnarray}\label{eq:Nn:N1}
    N_\pm^{(n)}(t)= \left(\frac{9}{8}\sqrt{6\pi}\frac{k_1}{k_2}\right)^{n-1}\widetilde{N}^{(1)}_\pm\left(\gamma_l(t)\left(\frac{128}{27}\right)^{n-1},t\right).
\end{eqnarray}
For an initial electron we simply have $F_\gamma^{(0)}(\gamma_\gamma,t)=I_\gamma(\gamma_\gamma,\gamma_0)$ and it follows from Eq. \eqref{eq:rec:NFn}:
\begin{eqnarray}\label{eq:lt:sol:N1}
    \widetilde{N}_\pm^{(1)}(\gamma_l,t)=\frac{9k_1}{2k_2}\left[\ln\left(\frac{\gamma_0}{\gamma_l(t)}\right)+3\left(\frac{\gamma_l(t)}{\gamma_0}\right)^{1/3}-3\right]\Theta [\gamma_0-\gamma_l(t)].
\end{eqnarray}
By using  $\chi_l(t)/\chi_0=\gamma_l(t)/\gamma_0$ and combining Eqs. \eqref{eq:Nn:N1} and \eqref{eq:lt:sol:N1} we have:
\begin{eqnarray}\label{eq:lt:Nn:approx1}
    N^{(n)}_\pm(t)\simeq c_3\times c_4^{n-1}\times 
     [\ln\left(\frac{\chi_0}{\chi_l(t)}\right) -c_6(n-1) \, +3c_7^{n-1}\left(\frac{\chi_l(t)}{\chi_0}\right)^{1/3}-3 ]\Theta \left[\chi_0-\chi_l(t)c_5^{n-1}\right].
\end{eqnarray}
with  $c_3\simeq3^{5/3}2.44^{2/3}\Gamma(-1/3)/(2\Gamma(-2/3)\Gamma(1/3))\simeq 4.22$, $c_4=3/4\sqrt{2\pi/3}c_3\simeq4.58$, $c_5=128/27 \simeq 4.74$, $c_6=\ln(128/27)\simeq 1.56$ and $c_7=(128/27)^{1/3}\simeq 1.68$. 

This equation allows to find a solution for the maximum number of generations, $n_{\rm\scriptscriptstyle{max}}$, in the shower. Since $N_\pm^{(n)}(t)$ is non zero only if $c_5^{n-1}<\chi_0/\chi_l(t)$ we find:
\begin{eqnarray}
    n_{\rm\scriptscriptstyle{max}}=1+c_6^{-1}\ln\left(\frac{\chi_0}{\chi_l(t)}\right)\ln^{-1}.
\end{eqnarray}
We are now in a position to obtain the shower multiplicity created by $N_0$ monoenergetic electron initially at energy $(\gamma_0-1)mc^2$ by summing the solution Eq. \eqref{eq:lt:Nn:approx1} over all accessible generation:
\begin{eqnarray}\label{eq:lt:Ntot:sumgen}
    N_\pm(t)/N_0&=&\sum_{i=1}^{n_{\rm\scriptscriptstyle{max}}}N_\pm^{(n)}(t).
\end{eqnarray}
In the limit $\chi_l \ll \chi_0$ we obtain the asymptotical expression for the shower multiplicity:
\begin{eqnarray}\label{eq:lt:Nn:approx2}
    N_\pm(t)/N_0\simeq c_8\chi_0\ln{\left(c_2 \chi_0^{1/3}\frac{t}{T_r}\right)}
\end{eqnarray}
with $c_8=\frac{9}{8}c_3\left(\frac{c_4c_7}{c_4c_7-1}+\frac{c_4c_6}{3(c_4-1)^2}-\frac{c_4}{c_4-1}\right)\simeq 0.26$.

\section{Application to beam-beam scattering}

We consider the collision of two uniform beams of electrons at energy $\mathcal{E}_0$ (Lorentz factor $\gamma_0$), density $n_0$ and spatially contained in a cylinder of radius $R$ and length $L$. We used the cylindrical coordinates. The field of one electron beam in its referential is perpendicular to its propagation direction and is given by:
\begin{eqnarray}
    E_\perp=-\frac{en_0}{2\epsilon_0}r
\end{eqnarray}
for $r\leq R$. A magnetic field is also present, perpendicular to the electric field and its value is given by $B_\perp=\sqrt{1-1/\gamma_0}E_\perp\simeq E_\perp$ for high energetic beams. It follows that for a perfect head-on collision, the quantum parameters of one electron of energy $mc^2(\gamma_0-1)$ located at in $r$ is:
\begin{eqnarray}
\chi(r)=\chi_0\frac{r}{R}
\end{eqnarray}
with
\begin{eqnarray}
\chi_0=2\gamma_0\frac{E_\perp(R)}{E_s}\simeq 26.8 \times \left(\frac{\mathcal{E}_0}{10~{\rm GeV}}\right)\,\left(\frac{n_0}{10^{24} {\rm cm}^{-3}}\right)\,
\left(\frac{R}{0.1 {\rm \mu m}}\right).
\end{eqnarray}
From this simple analysis, it is possible to calculate the analogue radiation time and $B/B_s$ in order to use the multiplicities of this work. The radiation time is define as $T_r(r)=c_1\tau_c/\alpha \, \gamma_0\chi(r)^{-2/3}$ so the minimal (at $r=R$) radiation time is:
\begin{eqnarray}
    T_r(R)\simeq 3.12  \,\, \rm{ fs} \, \left(\frac{\mathcal{E}_0}{10 \, \rm{GeV}}\right)^{\frac{1}{3}} \left(\frac{n_0}{10^{24} \, \rm{cm}^{-3}}\right)^{-\frac{2}{3}} \left(\frac{R}{0.1 \, \mu\rm{m}}\right)^{-\frac{2}{3}}.
\end{eqnarray}
For a beam of length $L$ the interaction time is $L/2/c$. For $L=1\mu$m it corresponds to an interaction time of $1.7$ fs and it follows that the short-time solutions Eqs. \eqref{eq:sh:sol:fpn}, \eqref{eq:sh:sol:fgn} and \eqref{eq:st:N1exact} can be used for a beam with laser-wakefield parameters ($n_0\simeq10^{22}$ cm$^{-3}$, $\mathcal{E}_0\simeq 10 $ GeV, $R\simeq1 \mu$m) . 

In cylindrical coordinates \cite{pouyez2024multiplicity}, the number of pairs generated by one beam is:
\begin{eqnarray}
    N_\pm(t)=n_0\int_0^{2\pi}d\theta  \int_0^{R}rdr\int_{0}^{L} dz M(r,z,t,\gamma_0).
\end{eqnarray}
The multiplicity $M(r,z,t,\gamma_0)$ is calculated using Eq. \eqref{eq:st:N1exact} for the first generation and corresponds to the number of pairs produced at a time $t$ by a single electron initially at $z$. Considering a time for which the interaction is over, all electrons have interacted during a time $L/2/c$ so in a regime $\chi_0\lesssim 1$, $M(r,z,t,\gamma_0)=[L/(2cT_r(r))]^2\times 1.25\chi(r)^{2/3}e^{-\frac{16}{3\chi(r)}}$. It follows that the final multiplicity reads:
\begin{eqnarray}\label{eq:beambeam:N}
    N_\pm/N_0\simeq c_9\,\left(\frac{L}{c\,T_r(R)}\right)^2\,\chi_0^{-10/3}\,\Gamma\left(-4,\frac{16}{3\chi_0}\right)\,,
\end{eqnarray}
with $c_9=1.25(16/3)^4\simeq 1.01\cdot10^{3}$, and $\Gamma$ the upper incomplete gamma function.
For cases where $\chi_0\gg1$ the same integration should be perform using the multiplicity $M(r,z,t,\gamma_0)=(t/T_r)^2\times (6.1\ln(\chi(r))-13.6)$.

\section{Application to laser scattering experiments}

 \begin{figure}[h]
    \centering
    \includegraphics[width=0.8\linewidth]{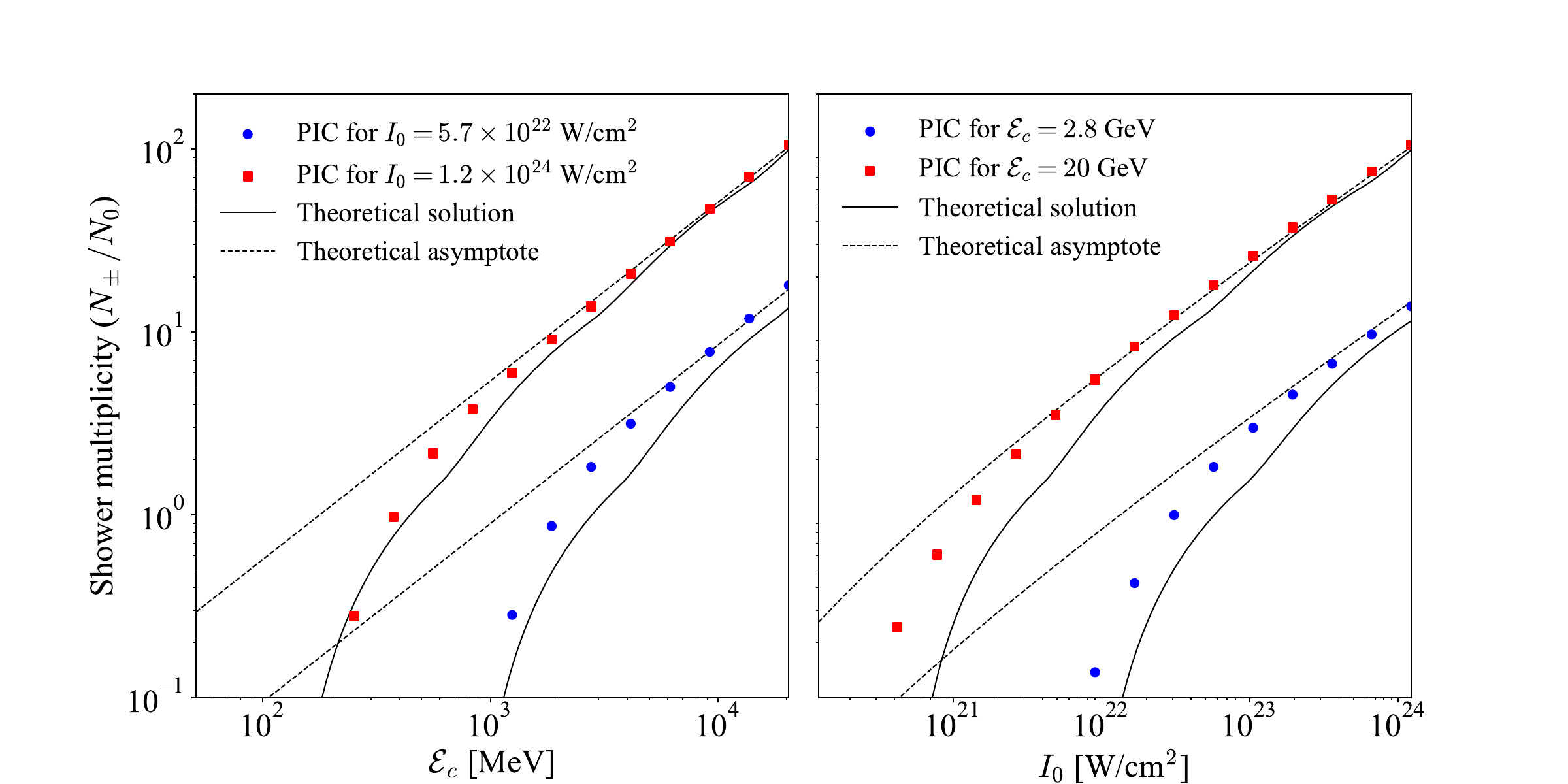}
    \caption{Shower multiplicity in typical laser scattering experiments with a laser-like-ELI ($T_{\rm\scriptscriptstyle{L}}=150$ fs and $\lambda=1.057 \, \mu$m). In the left panel the multiplicity as a function of the incident electron's energy for two laser intensity $I_0=5.7\times10^{22}$ W/cm$^2$ and $I_0=1.2\times10^{24}$ W/cm$^2$. In the right panel the multiplicity as a function of the laser intensity for two incident electron's energy $2.8$ GeV and $20$ GeV. Blue and red points are extracted from 1D3V PIC simulations, solid black line corresponds to Eq. \eqref{eq:lt:Ntot:sumgen} while the dashed black line stands for Eq. \eqref{eq:lt:Nn:approx2}.}
    \label{fig:scal_vs_eli}
\end{figure}

We consider the head-on collision of electrons beam at energy $\mathcal{E}_0=(\gamma_0-1)mc^2$ with a linearly polarized laser. The electric field profile of the laser is set to be:
\begin{eqnarray}\label{eq:Elaser}
    \vE(z,t) = E_0\,\sin(\omega\,(t-z/c))\,\sin\left(\frac{\pi}{2}\frac{t-z/c}{T_{\rm\scriptscriptstyle{L}}}\right)\,{\bf {\hat y}}\quad{\rm for}\,\, 0\le t-z/c < 2T_{\rm\scriptscriptstyle{L}}\,\,{\rm and}\,\,0\,\,{\rm otherwise}\,,
\end{eqnarray}
with $T_{\rm\scriptscriptstyle{L}}$ the full-width at half-maximum in intensity duration, $\omega=2\pi c/\lambda$ the pulsation and $E_0$ the maximal electric field corresponding to an intensity $I_0$.
Considering $T_{\rm\scriptscriptstyle{L}}\gg \lambda/c$ and that particles penetrate through the laser field at a velocity $c$, the average field seen by a particle is:
\begin{eqnarray}\label{eq:AvElaser}
    \langle E \rangle \simeq \frac{4}{\pi^2} E_0
\end{eqnarray}
during a time $T_{\rm\scriptscriptstyle{L}}$. Considering the average electric field, the equivalent quantum parameter read:
\begin{eqnarray}\label{eq:chi0laser}
    \chi_0&\simeq& 2\gamma_0\frac{\langle E\rangle}{E_0} = 5.17 \left( \frac{\mathcal{E}_0}{10\,  \rm{ GeV}}\right)\left(\frac{I_0}{10^{22} \, \rm{ W/cm}^2}\right)^{1/2}
\end{eqnarray}
and the equivalent radiation time:
\begin{eqnarray}\label{eq:Trlaser} 
    T_r= c_1\alpha^{-1}\,\tau_c \, \gamma_0\chi(r)^{-2/3}=9.35 \, \rm{ fs} \times \left( \frac{\mathcal{E}_0}{10 \rm{ GeV}}\right)^{1/3}\left(\frac{I_0}{10^{22} \, \rm{ W/cm}^2}\right)^{-1/3}
\end{eqnarray}

Considering pulses as the ones delivered by the L4 ATON laser at ELI Beamlines, we have $T_{\rm\scriptscriptstyle{L}}\simeq 150 {\rm fs}\gg \lambda/c$ and $T_{\rm\scriptscriptstyle{L}}\simeq 150 {\rm fs} \gg T_r$ for a wide range of parameters so that one can compute the final shower multiplicity using the long-time-limit Eq.~\eqref{eq:lt:Nn:approx2} at time $t=T_{\rm\scriptscriptstyle{L}}$. In Fig \ref{fig:scal_vs_eli} we show the shower multiplicity in typical laser scattering experiments with a laser of $T_{\rm\scriptscriptstyle{L}}=150$ fs and $\lambda=1.057 \, \mu$m. Red and blue points are obtained from 1D3V PIC simulations while black solid lines represent Eq. \eqref{eq:lt:Ntot:sumgen} and the dashed line corresponds to Eq. \eqref{eq:lt:Nn:approx2}. As shown, there is excellent agreement on the shape but also the order of magnitude between full 1D3V PIC simulations and the long-time solution for all the probed parameters. The differences mainly occur at low multiplicity regimes and are due to the consideration of an average field Eq. \eqref{eq:AvElaser}, even if Eq. \eqref{eq:lt:Ntot:sumgen} follow more closely the trend. As expected at low multiplicity the processes of pairs and photon emission are rare and the shape of the field strongly impacts their occurrence as opposed to when the multiplicity is sufficiently high and when a lot of particles are created quasi-everywhere. It is also worth mentioning that the solution Eq. \eqref{eq:lt:Ntot:sumgen} shows slow oscillation due to the contribution of a new generation.